%% file: 1570182913.tex
\documentclass[10pt,final,journal,twocolumn]{IEEEtran}
\usepackage[utf8]{inputenc}
\usepackage{graphicx}
\usepackage{subfigure}
\usepackage{amsmath}
\usepackage{amssymb}
\usepackage{color}
\usepackage{multirow}
\usepackage{algorithm}
\usepackage{algpseudocode}
\usepackage{tikz}
\usepackage[colorlinks,bookmarksopen,bookmarksnumbered,citecolor=red,urlcolor=red]{hyperref}

\newcommand\copyrighttext{%
  \footnotesize \textcopyright 2015 IEEE. Personal use of this material is permitted.
  Permission from IEEE must be obtained for all other uses, in any current or future 
  media, including reprinting/republishing this material for advertising or promotional 
  purposes, creating new collective works, for resale or redistribution to servers or 
  lists, or reuse of any copyrighted component of this work in other works. 
  DOI: \href{http://dx.doi.org/10.1109/JSAC.2015.2478996}{10.1109/JSAC.2015.2478996}
}
\newcommand\copyrightnotice{%
\begin{tikzpicture}[remember picture,overlay]
\node[anchor=south,yshift=10pt] at (current page.south) {\fbox{\parbox{\dimexpr\textwidth-\fboxsep-\fboxrule\relax}{\copyrighttext}}};
\end{tikzpicture}%
}

\begin{document}

\title{Adaptive DRX Scheme to Improve Energy Efficiency in LTE Networks with Bounded Delay}

\author{Sergio~Herrería-Alonso, 
  Miguel~Rodríguez-Pérez,~\IEEEmembership{Member,~IEEE,} \\
  Manuel Fernández-Veiga,~\IEEEmembership{Senior Member,~IEEE,}
  and~Cándido~López-García%
  \thanks{The authors are with the Telematics Engineering Dept.,
    Univ. of Vigo, 36310 Vigo, Spain. Tel.: +34~986~813458;
    fax: +34~986~812116; email: sha@det.uvigo.es
    (S. Herrería-Alonso).}} 

\maketitle
\copyrightnotice

\begin{abstract}
  The Discontinuous Reception (DRX) mechanism is commonly employed in
  current LTE~networks to improve energy efficiency of user equipment
  (UE). DRX allows UEs to monitor the physical downlink control
  channel (PDCCH) discontinuously when there is no downlink traffic
  for them, thus reducing their energy consumption. However, DRX power
  savings are achieved at the expense of some increase in packet delay
  since downlink traffic transmission must be deferred until the UEs
  resume listening to the PDCCH. In this paper, we present a promising
  mechanism that reduces energy consumption of UEs using DRX while
  simultaneously maintaining average packet delay around a desired
  target. Furthermore, our proposal is able to achieve significant
  power savings without either increasing signaling overhead or
  requiring any changes to deployed wireless protocols.
\end{abstract}


\section{Introduction}
\label{sec:introduction}

Current generation wireless networks such as Long Term Evolution (LTE)
and LTE-Advanced (LTE-A) are able to achieve high data rates up to
$1\,$Gb/s by adopting several advanced modulation, coding and multiple
antenna techniques~\cite{3gpp_ts_36300}. This great increment on the
offered capacity for data transmission has also increased the power
demands of mobile devices substantially.

To improve user equipment (UE) battery lifetime, LTE supports
Discontinuous Reception (DRX)~\cite{3gpp_ts_36321,bontu09:drx} in both
the RRC\_IDLE and the RRC\_CONNECTED radio resource control (RRC)
states~\cite{3gpp_ts_36331}. DRX allows UEs which are not receiving
data from their corresponding eNodeB (eNB) to monitor the physical
downlink control channel (PDCCH) discontinuously. When UEs are not
listening to the PDCCH, they can enter a power saving mode in which
most of their circuits can be turned off, thus reducing power
consumption significantly. With DRX, the UE only wakes up periodically
to listen to the PDCCH for a while, returning to the low power mode if
no packet arrival is detected or resuming its normal operation in the
case of new packet arrivals.

Obviously, DRX power savings are achieved at the expense of increasing
packet delay since all the traffic for UEs in the low power mode must
be buffered at the eNB until they listen to the PDCCH
again. Therefore, a careful configuration of the main DRX parameters
is critical to maintain a reasonable latency for active traffic while
obtaining significant power savings at the
UEs~\cite{bontu09:drx,fowler12:adjust_fix_drx}. DRX is configured per
UE (as opposed to per radio bearer) and there is only one DRX
configuration active in each UE at any time. Unfortunately, a single
DRX configuration fails to provide a satisfactory service to any
possible user activity level. Thus, several algorithms have been
proposed to configure DRX parameters according to the ongoing traffic
activity in a way that leads to a good trade-off between power savings
and packet
delay~\cite{bontu09:drx,karthik13:pr_alg_drx,koc14:drx_config,wen12:perf_drx,yu12:adj_drx,alouf12:power_saving_analysis,wang14:model_drx,wang14:drx_aware,tseng15:model_drx}. All
these schemes require, therefore, the reconfiguration of DRX
parameters whenever traffic characteristics change
substantially. However, DRX~reconfiguration is carried out via
RRC~signaling, so these schemes would cause considerable increments in
signaling overhead, especially when the number of UEs using DRX in a
given cell is large~\cite{3gpp_tr_25913}.

In this paper we present a promising mechanism to improve energy
efficiency of UEs using DRX that is able to control the average packet
delay with any fixed configuration of DRX parameters and, therefore,
does not increase signaling overhead. Our proposal is based on the
well-known packet coalescing technique successfully applied beforehand
to reduce energy consumption in other networking fields such as
Ethernet interfaces~\cite{christensen10:road_eee,herreria12:gig1} and
EPON systems~\cite{rodriguez12:up_epon,herreria14:doze_mode}. With
current DRX, an eNB with some downlink traffic queued for a given UE
in the low power mode will not send this traffic until the next time
the UE monitors the PDCCH. Notice that, as soon as the UE detects a
new packet arrival, it abandons DRX mode and returns to normal
operation. Therefore, the amount of time that UEs spend in the low
power mode can be easily increased if the eNB just delays packet
transmission to those UEs in DRX mode until their corresponding
downstream queues reach a certain threshold. However, a single
threshold value does not suit well for any possible downlink traffic,
so our proposal includes an adaptive algorithm able to adjust this
parameter to real time traffic characteristics with the goal of
maintaining average packet delay around a given target while keeping
energy consumption low enough.

In addition, our proposal is very simple to setup since it only
requires configuring two straightforward parameters: the average and
the maximum queueing delay desired for packets at the
eNB. Furthermore, the algorithm can be easily deployed in current
LTE/LTE-A~networks since it only demands some minor changes to the
operations of the eNB, keeping existent wireless protocols unmodified.

\begin{figure*}[t]
  \centering
  \resizebox{0.95\textwidth}{!}{\input{drx-example.pstex_t}}
  \caption{DRX operations with $N_{\mathrm{s}}=3$.}
  \label{fig:drx_example}
\end{figure*}
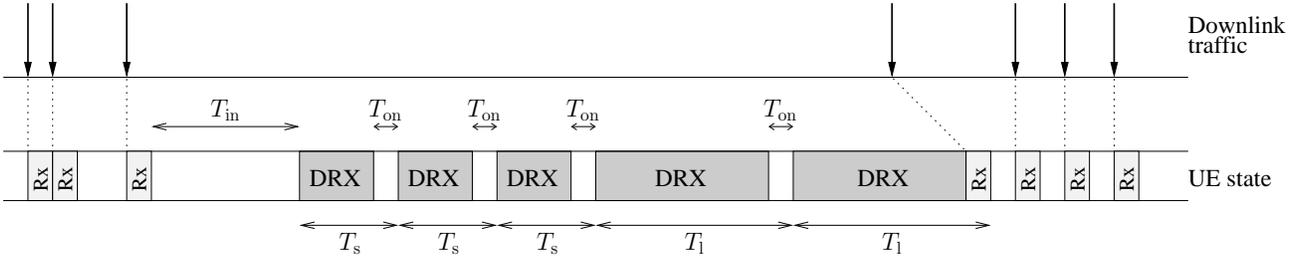

The rest of this paper is organized as follows. Section~\ref{sec:drx}
presents the basic operations of DRX and
Section~\ref{sec:coalesced-drx} describes the proposed scheme. We
refer to our proposal as coalesced DRX. In
Section~\ref{sec:delay-model} we develop an analytic model to compute
the average packet delay introduced by coalesced DRX. We next
particularize this model to Poisson traffic in
Section~\ref{sec:poisson}. This model is used in
Section~\ref{sec:adaptive} to devise an adaptive algorithm able to
adjust the queue threshold used in coalesced DRX according to traffic
conditions. Section~\ref{sec:results} shows some results obtained
through simulation. In Section~\ref{sec:related} we discuss some
implementation issues and compare our proposal with previous work in
this field. Finally, the main conclusions are summarized in
Section~\ref{sec:conclusions}.

\section{DRX Operations}
\label{sec:drx}

The DRX mechanism is frequently used in LTE/LTE-A networks to reduce
the power consumption of mobile handsets. Although this mechanism can
be configured in both the RRC\_IDLE and the RRC\_CONNECTED states, we
assume that UEs are always in the latter state, as it is the usual case for
those UEs with multiple applications running in the
background~\cite{gupta13:energ_impact}.

In the RRC\_CONNECTED state, a two-level power-saving scheme with both
short and long DRX~cycles is used. Figure~\ref{fig:drx_example}
depicts a typical example of the main DRX~operations. When DRX is
enabled, UEs stop listening to the PDCCH and enter a low power
mode. While in this sleeping mode, UEs cannot receive packets, so the
eNB must delay the transmission of all their downlink traffic until
they monitor the PDCCH again. Then, sleeping UEs periodically wake up
to listen to the PDCCH for a short interval to check for new packet
arrivals.

DRX configuration involves setting various parameters during the radio
bearer establishment. The DRX~parameters considered in this paper are:
\begin{itemize}
\item Inactivity timer ($T_{\mathrm{in}}$): time to wait before
  enabling DRX. This timer is immediately re-initiated after a
  successful reception on the PDCCH. When this timer expires, the UE
  enables DRX and enters the short DRX cycle.
\item Short DRX cycle ($T_{\mathrm{s}}$): duration of the first DRX
  cycles after enabling DRX.
\item DRX short cycle timer ($N_{\mathrm{s}}$): long DRX cycles will
  be applied after this timer expires. It is usually expressed as the
  number of short DRX cycles before transitioning to long DRX cycles.
\item Long DRX cycle ($T_{\mathrm{l}}$): duration of DRX cycles after
  $N_{\mathrm{s}}$ short DRX cycles ($T_{\mathrm{l}} \ge T_{\mathrm{s}}$).
\item On-duration timer ($T_{\mathrm{on}}$): interval at every DRX
  cycle during which the UE monitors the PDCCH checking for the
  arrival of a new packet ($T_{\mathrm{on}} \ll T_{\mathrm{s}}$). A
  successful reception on the PDCCH during this interval finishes the
  DRX cycle immediately and the inactivity timer is started again.
\end{itemize}

We also assume that the LTE~network is lightly loaded and that,
therefore, radio resources are always available to the UE when
required. It is expected that this will be the most likely scenario in
the near future since mobile networks are now rapidly evolving to
include small cells (such as picocells and femtocells), thus reducing
the number of users competing for resources at each base
station~\cite{andrews13:hetnets}. On the other hand, if the network
were lightly congested, a DRX-aware scheduling scheme would be
required at the eNB to resolve resource contention among UEs. This is
out of the scope of our paper, but a scheduler similar to those
proposed
in~\cite{bo10:drx_aware_scheduling,liang13:energy_efficient_scheduling}
that keeps aware of DRX operations and mitigates resource contention
could be applied.

\begin{figure*}[t]
  \centering
  \resizebox{0.95\textwidth}{!}{\input{coalescing-drx.pstex_t}}
  \caption{Coalesced DRX with $Q_{\mathrm{w}}=3\,$packets.}
  \label{fig:coalesced-drx}
\end{figure*}
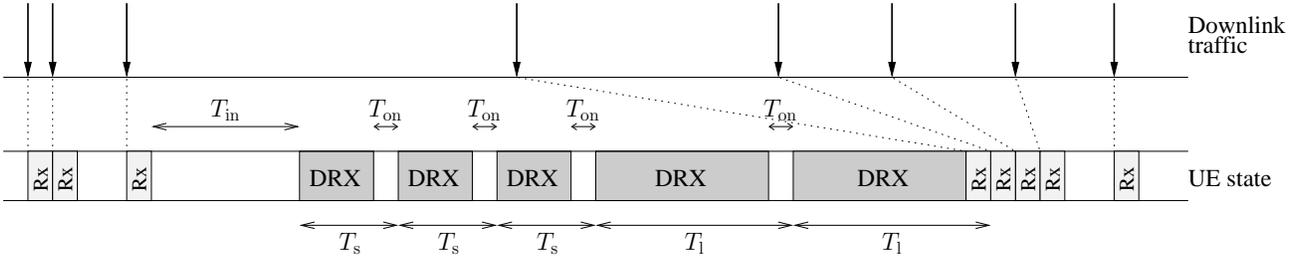

\section{Coalesced DRX}
\label{sec:coalesced-drx}

As explained in the previous section, an eNB with some downlink
traffic queued for a given UE in DRX mode will not send this traffic
until the UE monitors the PDCCH again (i.e., in the next on-duration
interval). Therefore, the eNB is already carrying out an implicit
packet coalescing because a significant period of time may be spanned
since it receives a packet for the sleeping UE until the actual
beginning of transmission ($T_{\mathrm{l}}-T_{\mathrm{on}}$ at
most). However, note that current eNBs will start downlink delivery
even if only a single packet is awaiting transmission.

In this paper we propose that eNBs delay downlink transmission to UEs
in DRX mode until their corresponding downstream queues reach a
certain tunable threshold
($Q_{\mathrm{w}}$). Figure~\ref{fig:coalesced-drx} shows an example of
coalesced DRX operations with
$Q_{\mathrm{w}}=3\,$packets.\footnote{Although the queue threshold is
  specified in packets for simplicity, in a real setting it should be
  specified in bytes to handle packets of different sizes.}  Instead
of starting packet transmission as soon as there is new traffic to
transmit, the eNB waits until $Q_{\mathrm{w}}$~packets are
accommodated in the downstream queue, thus reducing the number of
transitions between the DRX and the active mode and, therefore,
increasing the amount of time the UE remains in the low power
mode. Note that downlink transmission is delayed even when new packets
arrive during on-duration intervals if the $Q_{\mathrm{w}}$~threshold
has not yet been reached. Once the DRX mechanism has been disabled,
packets arriving before the inactivity timer expires are transmitted
as soon as possible. In short, what our proposal makes in practice is
just adjusting the duration of DRX cycles, but without the need for
performing an actual reconfiguration of DRX parameters.

\begin{figure*}[t]
  \centering
  \resizebox{0.9\textwidth}{!}{\input{drx-cycle.pstex_t}}
  \caption{Coalesced DRX cycle with $Q_{\mathrm{w}}=3\,$packets.}
  \label{fig:coalesced-drx-cycle}
\end{figure*}
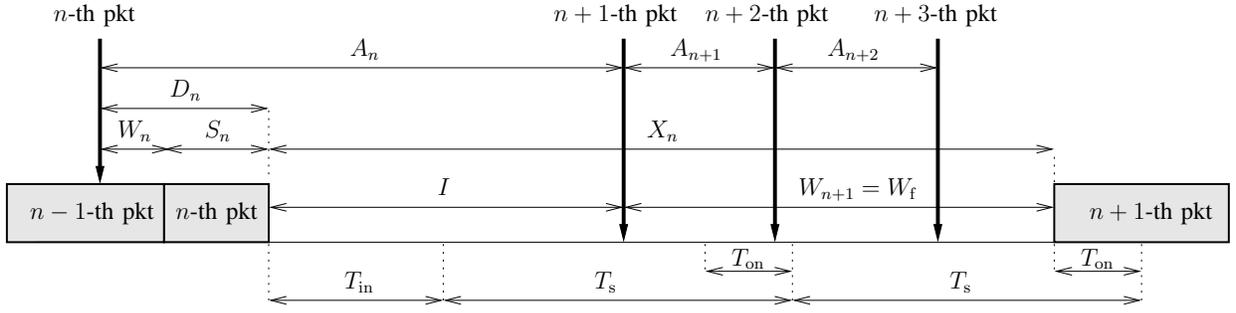

Obviously, this backlog will provide greater energy savings in the UEs
at the expense of increasing packet delay. To avoid delaying downlink
traffic excessively, the maximum time~$W_{\max}$ an eNB can delay
packet transmission must be bounded.

\section{Delay Model}
\label{sec:delay-model}

In this section, we will quantify the impact of coalesced DRX on
packet delay. Particularly, we compute the average queueing delay,
$\mathrm{E}[W]$, following a similar approach to that used
in~\cite{marshall68:bounds_gig1} for queueing systems with vacations
in which the first packet in every busy period suffers a random
delay~$W_{\mathrm{f}}$ before its transmission begins. For simplicity,
we assume that $T_{\mathrm{s}}=T_{\mathrm{l}}$, that is, all DRX
cycles are of equal length.\footnote{This is the default configuration
  in actual DRX systems since short DRX cycles are optional and need
  to be explicitly configured. In any case, considering different long
  and short DRX cycles would only obfuscate the theoretical analysis
  without providing further insight into the operations of our
  proposal since it does not rely on any particular DRX
  configuration.}

Figure~\ref{fig:coalesced-drx-cycle} illustrates several definitions
that will be used in the model. We consider that packet arrivals at
the eNB for a given UE follow a general distribution with independent
inter-arrival times $A_n$, $n=1,2,\ldots$, and average arrival
rate~$\lambda$. The sequence of service times $S_n$, $n=1,2,\ldots$,
demanded by successive packets is a set of random variables with a
common, although arbitrary, distribution function and mean service
rate~$\mu$. Obviously, the utilization factor $\rho=\lambda/\mu$ must
be less than~$1$ to assure system stability. A notation summary is
provided in Table~\ref{tab:notation}.

\begin{table*}[t]
  \caption{Notation summary.}
  \label{tab:notation}
  \begin{center}
    \small
    \begin{tabular}{|c|c|c|}
      \hline
      \textbf{Category} & \textbf{Notation} & \textbf{Description} \\
      \hline
      \multirow{5}{*}{Traffic characteristics} & $A_n$ & Time between the arrivals of $n$-th and $(n+1)$-th packets \\
      & $S_n$ & Service time demanded by $n$-th packet \\
      & $\lambda$ & Average arrival rate \\
      & $\mu$ & Average service rate \\
      & $\rho$ & Utilization factor \\
      \hline
      \multirow{9}{*}{Model variables} & $O_n$ & Time between the end of transmissions of $n$-th and $(n+1)$-th packets \\
      & $D_n$ & Total time the $n$-th~packet spends in the eNB \\
      & $W_n$ & Queueing delay of the $n$-th~packet \\
      & $X_n$ & Idle time since end of $n$-th packet tx until start of $(n+1)$-th packet tx \\
      & $Y_n$ & $X_n(X_n+2(D_n-A_n))$ \\
      & $U_n$ & $S_n-A_n$ \\
      & $I$ & Length of the empty period in the coalescing cycle \\
      & $W_{\mathrm{f}}$ & Queueing delay of the first coalesced packet in the cycle \\
      & $\gamma$ & Inverse of the fraction of idle time at the eNB with DRX enabled \\
      \hline
      \multirow{5}{*}{DRX parameters} & $T_{\mathrm{in}}$ & Inactivity timer \\
      & $T_{\mathrm{on}}$ & On-duration timer \\
      & $T_{\mathrm{s}}$ & Short DRX cycle duration \\
      & $T_{\mathrm{l}}$ & Long DRX cycle duration \\
      & $N_{\mathrm{s}}$ & DRX short cycle timer \\
      \hline
      \multirow{4}{*}{Coalesced DRX parameters} & $Q_{\mathrm{w}}$ & Queue threshold \\
      & $Q_{\max}$ & Maximum queue threshold \\
      & $W^*$ & Target average queueing delay \\
      & $W_{\max}$ & Maximum queueing delay \\
      \hline
    \end{tabular}
  \end{center}
\end{table*}

\subsection{Average Queueing Delay}

Let $O_n$ be the inter-output times, that is, the time between the end
of transmissions of $n$-th and $(n+1)$-th packets. Clearly,
\begin{equation}
  \label{eq:O1}
  O_n=A_n+D_{n+1}-D_n,
\end{equation}
where $D_n=W_n+S_n$ is the total time that the $n$-th~packet spends in
the~eNB and encompasses the corresponding queueing and service times
($W_n$ and $S_n$, respectively). On the other hand, inter-output times
can be also obtained as:
\begin{equation}
  \label{eq:O2}
  O_n=X_n+S_{n+1},
\end{equation}
where $X_n$ is the idle time elapsed since the end of the $n$-th
packet transmission until the beginning of the $(n+1)$-th packet
transmission, so equating~\eqref{eq:O1} and~\eqref{eq:O2} we get
\begin{equation}
  \label{eq:W1}
  W_{n+1}=X_n+D_n-A_n.
\end{equation}
Squaring both sides of~\eqref{eq:W1}, we have
\begin{IEEEeqnarray}{rCl}
  W^2_{n+1} &=& X_n(X_n+2(D_n-A_n)) + (A_n-D_n)^2 \nonumber \\ 
  &=& Y_n + (S_n-A_n)^2 + W^2_n + 2W_n(S_n-A_n),
\end{IEEEeqnarray}
where $Y_n\triangleq X_n(X_n+2(D_n-A_n))$. Then, taking expectations
and assuming stationarity in the system, it follows that
\begin{equation}
  \label{eq:W2}
  \mathrm{E}[W] = \frac{-\mathrm{E}[U^2] - \mathrm{E}[Y] + 
    2\sigma(W,A)}{2\mathrm{E}[U]},
\end{equation}
where $U_n\triangleq S_n-A_n$ and $\sigma(W,A)\triangleq
\sigma(W_n,A_n)$ for packet~$n$. Although the subscript $n$ is
suppressed for clarity, it should be kept in mind that $\sigma(W,A)$
is the covariance between the wait in queue of some packet and the
time until the next packet arrives.
To calculate $\mathrm{E}[Y]=\mathrm{E}[X_n(X_n+2(D_n-A_n))]$, we must
consider the following three cases:
\begin{description}
\item[Case 1)] \ \ $A_n-D_n \le 0$: this occurs when a new packet
  arrives at the eNB while it is transmitting traffic to the UE, so
  $X_n=0$.
\item[Case 2)] \ \ $0 < A_n-D_n \le T_{\mathrm{in}}$: a new packet
  arrives at the eNB before the inactivity timer expires, so DRX has
  not yet been enabled and the packet can be immediately
  transmitted. Therefore, $X_n=A_n-D_n=I$, where $I$ is the length of
  the empty period in the coalescing cycle, that is, the period with
  no packets in the downstream queue.
\item[Case 3)] \ \ $A_n-D_n > T_{\mathrm{in}}$: a new packet arrives
  at the eNB with DRX active, so it cannot be transmitted until the UE
  re-listens to the PDCCH during a subsequent on-duration interval. In
  this case, $X_n=I+W_{\mathrm{f}}$, where $W_{\mathrm{f}}$ is the
  time that the first arriving packet in the coalescing cycle has to
  wait before it can be transmitted, that is, the queueing delay
  experienced by the first coalesced packet.
\end{description}
If we denote by $p_k, k \in \{1,2,3\}$, the probability that a new
packet arrives at the eNB under the conditions determined by former
case~$k$, then
\begin{IEEEeqnarray}{rCl}
  \label{eq:Y}
  \mathrm{E}[Y] &=& p_2 \mathrm{E}[I(I-2I)] + p_3 \mathrm{E}[(I+W_{\mathrm{f}})(I+W_{\mathrm{f}}-2I)] \nonumber \\ 
  &=& p_3 \mathrm{E}[W_{\mathrm{f}}^2] - (p_2+p_3) \mathrm{E}[I^2] \nonumber \\ 
  &=& p_3 \mathrm{E}[W_{\mathrm{f}}^2] - (1-p_1) \mathrm{E}[I^2].
\end{IEEEeqnarray}

On the other hand, from~\eqref{eq:W1} we know that
\begin{equation}
  \label{eq:X}
  X_n = W_{n+1}-D_n+A_n = W_{n+1}-(W_n+S_n)+A_n,
\end{equation}
so 
\begin{equation}
  \label{eq:X2}
  \mathrm{E}[X]=\mathrm{E}[A]-\mathrm{E}[S]=-\mathrm{E}[U].
\end{equation}
Additionally, $\mathrm{E}[X]$ can be obtained as
\begin{equation}
  \label{eq:X3}
  \mathrm{E}[X] = p_2 \mathrm{E}[I] + p_3 \mathrm{E}[I+W_{\mathrm{f}}]
  = p_3 \mathrm{E}[W_{\mathrm{f}}] + (1-p_1) \mathrm{E}[I].
\end{equation}
Therefore, substituting~\eqref{eq:Y},~\eqref{eq:X2} and \eqref{eq:X3}
into~\eqref{eq:W2}, and replacing $(1-p_1)/p_3$ with $\gamma$, we get
\begin{equation}
  \label{eq:W3}
  \mathrm{E}[W] = -\frac{\mathrm{E}[U^2]}{2\mathrm{E}[U]} + \frac{\mathrm{E}[W_{\mathrm{f}}^2] -
    \gamma \mathrm{E}[I^2]}{2(\mathrm{E}[W_{\mathrm{f}}] + \gamma \mathrm{E}[I])} + 
  \frac{\sigma(W,A)}{\mathrm{E}[U]}.
\end{equation}
In the following subsection, we will explain in greater detail the
significance of the $\gamma$~factor introduced. We now calculate
the first two moments of variable~$U$ simply as
\begin{equation}
  \label{eq:U}
  \mathrm{E}[U] = \mathrm{E}[S-A] = \mathrm{E}[S]-\mathrm{E}[A] = 
  \frac{1}{\mu} - \frac{1}{\lambda} = -\frac{1-\rho}{\lambda},
\end{equation}
and
\begin{equation}
  \label{eq:U2}
  \mathrm{E}[U^2] = \sigma_U^2 + \mathrm{E}[U]^2 = \sigma_S^2 + \sigma_A^2 + 
  \left(\frac{1-\rho}{\lambda} \right)^2,
\end{equation}
since $\sigma(S_n,A_n) = 0$ if it is assumed that packet lengths are
independent from the arrival process.

Regarding $\sigma(W,A)$, note that the waiting time of the first
$Q_{\mathrm{w}}-1$ packets in each coalescing cycle depends on the
inter-arrival times of subsequent packets, so this covariance term must
be nonzero. In~\cite{heyman68:bounds_single_server_queues} it is
proved that, for single-server queues that wait until $Q_{\mathrm{w}}$
customers are present before starting service again, this covariance
term is given by
\begin{equation}
  \label{eq:cov_W_A}
  \sigma(W,A) = \frac{(1 - \rho) (Q_{\mathrm{w}} - 1)
    \sigma_A^2}{Q_{\mathrm{w}} - 1 + \lambda \mathrm{E}[I]}.
\end{equation}
Finally, substituting \eqref{eq:U}, \eqref{eq:U2} and
\eqref{eq:cov_W_A} into \eqref{eq:W3}, we get
\begin{IEEEeqnarray}{rCl}
  \label{eq:W4}
  \mathrm{E}[W] &=& \frac{\lambda^2(\sigma_S^2 + \sigma_A^2) + (1-\rho)^2}{2\lambda(1-\rho)} +
  \frac{\mathrm{E}[W_{\mathrm{f}}^2] - \gamma \mathrm{E}[I^2]}{2(\mathrm{E}[W_{\mathrm{f}}] + \gamma \mathrm{E}[I])} \nonumber \\
  & & - \frac{\lambda (Q_{\mathrm{w}}-1)\sigma_A^2}{Q_{\mathrm{w}} - 1 + \lambda \mathrm{E}[I]}.
\end{IEEEeqnarray}

\subsection{The $\gamma$ Factor}
\label{sec:gamma}

The $\gamma$ factor introduced in the previous analysis has been
defined as the ratio between the probabilities $1-p_1$ and
$p_3$. Recall that $p_1$ is the probability that a new packet arrives
at the eNB while it is transmitting downlink traffic to the UE, so
$1-p_1$ is the probability that a new packet arrives when the eNB is
idle. On the other hand, $p_3$ is the probability that a new packet
arrives when the eNB is idle but it has nevertheless to wait to be
served since DRX is active. So, $\gamma$ is the inverse of the
fraction of the idle time at the eNB with DRX enabled.

The probability $p_1$ can be obtained as 
\begin{equation}
  \label{eq:p1}
  p_1 = \mathrm{P}[A_n-D_n \le 0] = \mathrm{P}[A_n \le D_n] =
  \int_0^\infty F_A(t) f_D(t)\,\mathrm{d}t,
\end{equation}
where $F_A(t)$ is the cumulative distribution function of inter-arrival
times and $f_D(t)$ is the probability density function of the total
time spent by each packet in the eNB. Similarly, the probability $p_3$
is directly
\begin{equation}
  \label{eq:p3}
  p_3 = \mathrm{P}[A_n-D_n > T_{\mathrm{in}}] = 1 - \int_0^\infty
  F_A(t+T_{\mathrm{in}}) f_D(t)\,\mathrm{d}t.
\end{equation}
Therefore, $\gamma$ is
\begin{equation}
  \label{eq:gamma}
  \gamma = \frac{1 - \int_0^\infty F_A(t) f_D(t)\,\mathrm{d}t}{1 -
    \int_0^\infty F_A(t+T_{\mathrm{in}}) f_D(t)\,\mathrm{d}t} \ge 1.
\end{equation}

Notice from~\eqref{eq:W4} that the higher this factor is, the lower
packet delays we get. For example, if we configure DRX with a high
$T_{\mathrm{in}}$~value (a high $\gamma$~value), the added delay will
be reduced since the probability that a packet arrives at the eNB with
DRX disabled will be increased and, therefore, it will be more likely
transmitted at once. As usual, this reduction in packet latency can
only be obtained at the expense of increasing power consumption.

\section{Poisson Traffic}
\label{sec:poisson}

Here we will particularize the previous model for Poisson traffic,
that is, assuming that the number of packets that arrive at the eNB in
a given interval of time follows a Poisson distribution with average
arrival rate~$\lambda$ and variance $\sigma_A^2=1/\lambda^2$. Although
it is well-known that packet arrivals do not generally follow a
Poisson distribution, this approximation can be used to model
background traffic generated by mobile applications when the UE is in
unattended mode~\cite{koc14:drx_config}.

As can be seen from~\eqref{eq:W4}, to complete the computation of the
average queueing delay, we still have to calculate the average
duration of empty periods $\mathrm{E}[I]$ (and $\mathrm{E}[I^2]$), the
average waiting time of the first packet in each coalescing cycle
$\mathrm{E}[W_{\mathrm{f}}]$ (and $\mathrm{E}[W_{\mathrm{f}}^2]$), and
the $\gamma$ factor for this particular traffic distribution.

\subsection{Average Duration of Empty Periods}

With Poisson traffic, inter-arrival times are exponentially
distributed. Due to the memoryless property of this distribution, the
distribution of empty periods is equivalent to that of inter-arrival
times, so $\mathrm{E}[I]=1/\lambda$ and $\mathrm{E}[I^2]$ can be
easily calculated as $\mathrm{E}[I^2] = \mathrm{E}[A^2] =
\sigma_A^2+\mathrm{E}[A]^2 = 2/\lambda^2$.

\subsection{Average Waiting Time of the First Packet in Each Cycle}
\label{sec:poisson:wf}

With coalesced DRX, the first arriving packet in each coalescing cycle
has to wait for the arrival of other $Q_w-1$ packets before the eNB is
able to start its transmission, i.e., $(Q_{\mathrm{w}}-1)/\lambda$ on
average. Actually, transmission cannot begin until the destination UE
checks the PDCCH in the immediate on-duration interval following the
arrival of the $Q_w$-th packet. Consequently, the transmission of the
first backlogged packet will be delayed an extra time denoted by
$T_w$ and we have
\begin{equation}
  \label{eq:Wf}
  \mathrm{E}[W_{\mathrm{f}}] = \frac{Q_{\mathrm{w}}-1}{\lambda} + T_{\mathrm{w}},
\end{equation}
with
\begin{equation}
  \label{eq:Wf2}
  \mathrm{E}[W_{\mathrm{f}}^2] = \sigma_{W_{\mathrm{f}}}^2 + (\mathrm{E}[W_{\mathrm{f}}])^2
  = \frac{Q_{\mathrm{w}}-1}{\lambda^2} + \left( \frac{Q_{\mathrm{w}}-1}{\lambda} +
  T_{\mathrm{w}} \right)^2.
\end{equation}

To compute this additional delay $T_w$, we may distinguish two
different cases. First, if the $Q_w$-th packet (after entering the DRX
mode) arrives during an on-duration period, downlink transmission can
start at once and no extra delay is incurred. Conversely, if the
$Q_w$-th packet arrives with DRX active and the UE is not listening to
the PDCCH, a non-zero delay is added. Since Poisson arrivals are
independently and uniformly distributed on any interval of time, we
can assume that the arrival instant of the $Q_w$-th packet is
uniformly distributed along the DRX cycle interval, which gets more
true the higher the queue threshold or the inter-arrival times are,
and hence an average extra delay of
$(T_{\mathrm{s}}-T_{\mathrm{on}})/2$ will be introduced with
probability $(T_{\mathrm{s}}-T_{\mathrm{on}})/T_{\mathrm{s}}$. Therefore, on
average, $T_{\mathrm{w}} =
(T_{\mathrm{s}}-T_{\mathrm{on}})^2/(2T_{\mathrm{s}})$.

\subsection{The $\gamma$ Factor}

The $\gamma$ factor for Poisson traffic can be easily calculated just
substituting the cumulative distribution function of the exponential
distribution $F_A(t)=1-\mathrm{e}^{-\lambda t}, t \ge 0$,
into~\eqref{eq:gamma}:
\begin{equation}
  \label{eq:gamma_poisson}
  \gamma = \frac{1 - \int_0^\infty (1-\mathrm{e}^{-\lambda t}) f_D(t)\,\mathrm{d}t}{1 -
    \int_0^\infty (1-\mathrm{e}^{-\lambda (t+T_{\mathrm{in}})}) f_D(t)\,\mathrm{d}t}
  = \mathrm{e}^{\lambda T_{\mathrm{in}}}.
\end{equation}

\subsection{Average Queueing Delay}

Finally, substituting the results obtained for $\mathrm{E}[I]$ (and
$\mathrm{E}[I^2]$) and $\mathrm{E}[W_{\mathrm{f}}]$ (and
$\mathrm{E}[W_{\mathrm{f}}^2]$) into~\eqref{eq:W4}, the average
queueing delay with Poisson traffic can be estimated as
\begin{IEEEeqnarray}{rCl}
  \label{eq:W_poisson}
  \mathrm{E}[W] &=& \frac{1+\lambda^2\sigma_S^2+(1-\rho)^2}{2\lambda(1-\rho)} - \frac{Q_{\mathrm{w}}-1}{\lambda Q_{\mathrm{w}}} \nonumber \\
  & & + \frac{(Q_{\mathrm{w}}+\lambda T_{\mathrm{w}})^2 - Q_{\mathrm{w}} - 2(\lambda T_{\mathrm{w}} + \gamma)}{2\lambda(Q_{\mathrm{w}} + \lambda T_{\mathrm{w}} + \gamma -1)},
\end{IEEEeqnarray}
with
$T_{\mathrm{w}}=(T_{\mathrm{s}}-T_{\mathrm{on}})^2/(2T_{\mathrm{s}})$
and $\gamma=\mathrm{e}^{\lambda T_{\mathrm{in}}}$.

\section{Adaptive Coalesced DRX}
\label{sec:adaptive}

A good tuning of the queue threshold is key for the performance of the
coalesced DRX mechanism. If the queue threshold is too high, packets
can get excessively delayed. On the contrary, setting a too low
threshold reduces the power savings. An additional problem is that a
single threshold value does not suit well for any possible incoming
traffic. As shown later, when the traffic load is low, increasing the
threshold, even from modest values, produces unacceptable large
increments on packet delay with only marginal increments on power
savings. Under these circumstances, a low threshold is desirable, as
it provides small latencies with good enough energy savings. For high
traffic loads, the situation is just reversed. If the threshold were
not increased, power savings would be greatly diminished.

In this section we present an algorithm to dynamically accommodate the
$Q_{\mathrm{w}}$~parameter to incoming traffic. The main goal of our
algorithm is to minimize power consumption in UEs while trying to
maintain the average packet delay around a given target
value~$W^*$. So, to adjust the $Q_\mathrm{w}$ parameter to the
existent traffic conditions, the average delay experienced by packets
in a given coalescing cycle~$i$, that is, $\widehat{W}[i]$, should be
measured and compared with the target delay~$W^*$. Then, if
$\widehat{W}[i] > W^*$, $Q_\mathrm{w}$ should be reduced to diminish
packet delay. Conversely, if $\widehat{W}[i] \le W^*$, current average
packet delay is low enough and $Q_\mathrm{w}$ can be increased to
reduce power consumption.

We know, therefore, the direction in which the queue threshold should
be modified to make packet delay converge to the desired value, but we
still have to select a proper function to update this parameter
accurately. Intuitively, the more distant the measured average packet
delay is from the target value, the more aggressive changes in the
queue threshold should be. Therefore, we compute
from~\eqref{eq:W_poisson} the partial derivative of $\mathrm{E}[W]$
with respect to $Q_\mathrm{w}$ to understand how this parameter
affects average queueing delay in a given scenario:
\begin{equation}
  \label{eq:dW_dQ}
  \frac{\partial \mathrm{E}[W]}{\partial Q_{\mathrm{w}}} = \frac{1}{2\lambda} \left(1 - \frac{2}{Q_{\mathrm{w}}^2} + \frac{\lambda T_{\mathrm{w}}-\gamma(\gamma+1)}{(Q_{\mathrm{w}}+\lambda T_{\mathrm{w}}+\gamma-1)^2} \right).
\end{equation}
From this, it can be proved that $\partial \mathrm{E}[W]/\partial
Q_{\mathrm{w}} \le (2\lambda)^{-1}$ for all $Q_{\mathrm{w}} \ge 1$ if
$\lambda \le 4/T_{\mathrm{w}}$ or $\gamma \ge (\sqrt{4\lambda
  T_{\mathrm{w}}-7}-1)/2$, which holds true for usual DRX
parameters. Consequently, we propose to modify $Q_\mathrm{w}$ using a
conventional closed-loop controller with error signal
$W^*-\widehat{W}[i]$ and proportionality constant $2\lambda$ to obtain
a good compromise between $Q_{\mathrm{w}}$~stability and a fast
response to changing traffic conditions, as shown in
Algorithm~\ref{alg:tuning}. Also note that $Q_\mathrm{w}$ must not
exceed a maximum value $Q_{\max}=W_{\max}/S_{\max}$, with $S_{\max}$
being the maximum service time a packet could demand, to avoid
introducing queueing delays greater than~$W_{\max}$. The stability of
this dynamic algorithm is evaluated in the Appendix.

\begin{algorithm}[t]
  \caption{Tuning algorithm of $Q_\mathrm{w}$ executed at the end of
    each coalescing cycle, just before entering the DRX mode.}
  \label{alg:tuning}
  \begin{algorithmic}
    \Require Estimates of the average delay in current cycle $i$
    ($\widehat{W}[i]$) and the arrival rate ($\widehat{\lambda}$)
    \State $\displaystyle Q_\mathrm{w}[i+1]\gets Q_\mathrm{w}[i] +
    2\widehat{\lambda}(W^*-\widehat{W}[i])$
    \If{$Q_\mathrm{w}[i+1] < 1$}
    \State $Q_\mathrm{w}[i+1]\gets 1$
    \ElsIf{$Q_\mathrm{w}[i+1] > Q_{\max}$}
    \State $Q_\mathrm{w}[i+1]\gets Q_{\max}$
    \EndIf
  \end{algorithmic}
\end{algorithm}

To apply the proposed algorithm the eNB just needs to measure, for
each connected~UE, the average queueing delay in each coalescing cycle
and the arrival rate. As suggested
in~\cite{stoica03:core_stateless_fq}, we estimate the average arrival
rate~$\widehat{\lambda}$ using the following exponential moving
average:
\begin{equation}
  \widehat{\lambda}_n = \left( 1-\mathrm{e}^{-A_n/k} \right) \frac{1}{A_n} +
  \mathrm{e}^{-A_n/k} \widehat{\lambda}_{n-1},
\end{equation}
where $A_n$ is the time between the arrivals of $n+1$-th and $n$-th
packets and $k=2W_{\max}$. A variable weight $\mathrm{e}^{-A_n/k}$ is
used instead of a constant weight since, as stated
in~\cite{stoica03:core_stateless_fq}, this more closely reflects a
fluid averaging process independent of the packetizing structure.

\section{Simulation Results}
\label{sec:results}

To evaluate the performance of the proposed scheme, we conducted
several simulation experiments on an in-house simulator, available for
download at~\cite{herreria14:drx_simulator}. As performance metrics,
we select the power savings in the UE and the average queueing delay
experienced by downlink traffic due to DRX operations. To estimate
power savings, we measure the percentage of time spent by the UE in
the low power mode, thus avoiding the reliance on any particular power
consumption model.

Each simulation was run for $100\,$seconds and repeated ten times
using different random seeds. Then, an average of the measured
parameter was taken over all the runs. Although $95\,\%$ confidence
intervals have been also calculated, they will not be represented in
the graphs since all of them are small enough and just clutter the
figures.

In all the simulation experiments, the physical sub-frame (PSF) length
is set to $1\,$ms. We assume that each packet transmission requires
exactly one PSF.

\subsection{Coalesced DRX}

To evaluate coalesced DRX and validate our model, we consider Poisson
traffic with an increasing average arrival rate up to $0.9\,$packets
per PSF. In this experiment, we used the following conventional DRX
parameters: $T_{\mathrm{in}} = 10\,$ms, $T_{\mathrm{s}} =
T_{\mathrm{l}} = 32\,$ms and $T_{\mathrm{on}} = 2\,$ms.\footnote{The
  DRX cycle length has been selected to obtain a good trade-off
  between power efficiency and getting packet delays close to the
  target delay. Note that, with excessively short DRX cycles, the UE
  would check PDCCH too many times before reaching the queue threshold
  causing, therefore, an unnecessary high number of transitions
  between the DRX and the active mode that would increase power
  consumption. On the contrary, if DRX cycles were excessively long,
  the queue threshold could be reached well before the UE resumes
  listening to the PDCCH. This would prevent our adaptive proposal
  from achieving a good approximation to the target delay since the
  queue threshold could be only adjusted in a coarse-grained
  manner. We have checked through simulation that values between
  $16$--$64\,$ms for the DRX cycle length are suitable.}

\begin{figure}[t]
  \centering
  \subfigure[Energy savings.]{
    \includegraphics[width=0.95\columnwidth]{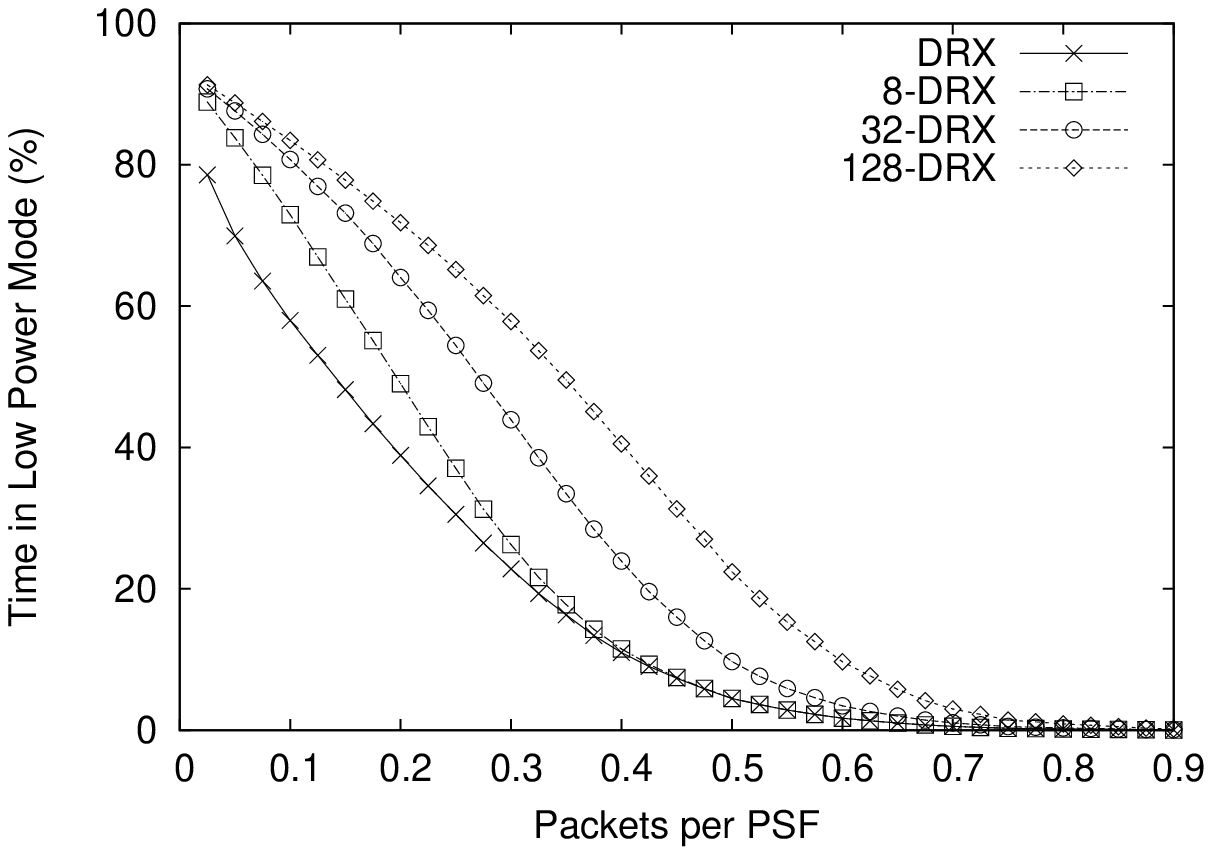}
    \label{fig:model-energy}
  } 
  \subfigure[Average queueing delay.]{
    \includegraphics[width=0.95\columnwidth]{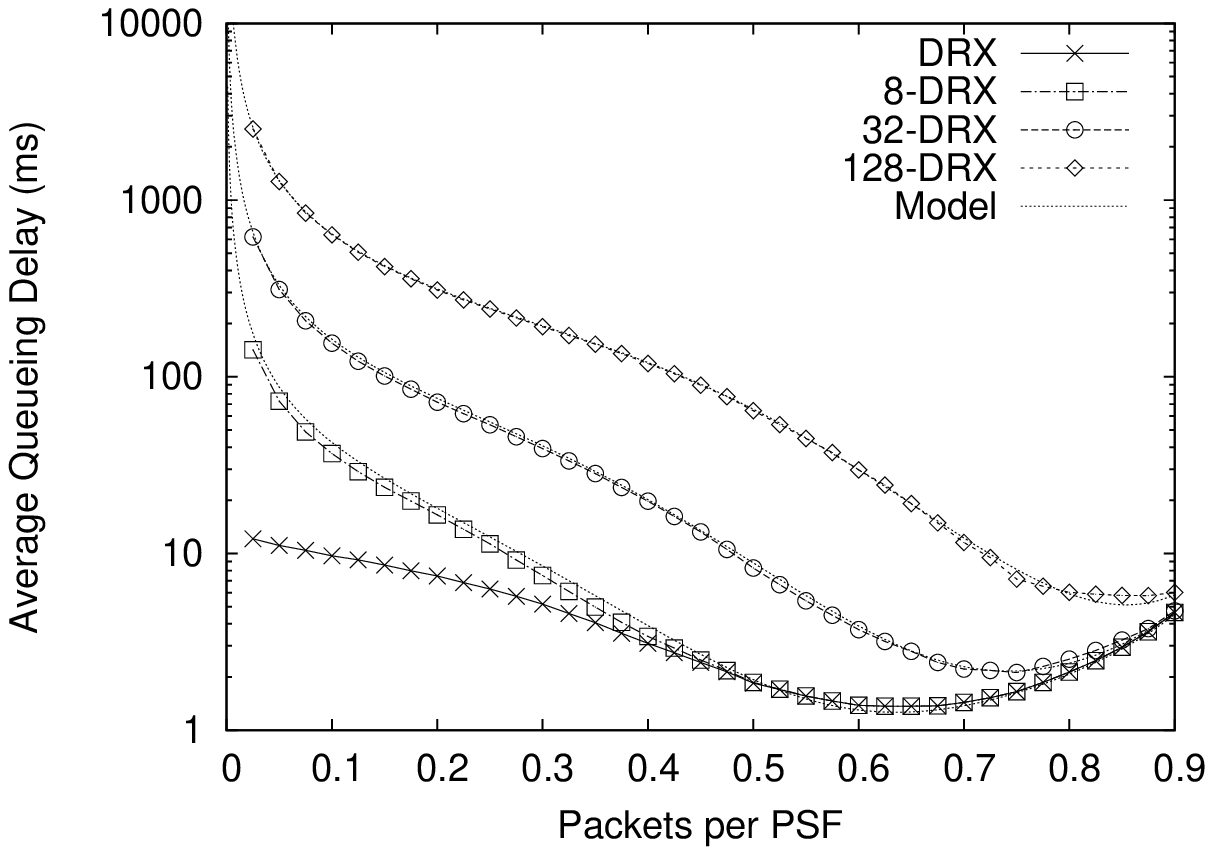}
    \label{fig:model-delay}
  }
  \caption{Results with coalesced DRX.}
  \label{fig:model}
\end{figure}

Figure~\ref{fig:model} shows the percentage of time spent in the low
power mode and the average queueing delay with conventional DRX (i.e.,
without coalescing) and when coalescing is applied with three
different queue thresholds ($Q_{\mathrm{w}} \in
\{8,32,128\}\,$packets). As expected, packet coalescing increases the
time spent in the low power mode at the expense of increasing packet
latency. Also note that our model produces very accurate predictions
for the average queueing delay in all the simulated scenarios.

Obviously, the higher the queue threshold is, the greater energy
savings are obtained but, at low rates, using a high $Q_{\mathrm{w}}$
leads to unacceptable large increments on packet delay to just achieve
marginal increments in power savings. Therefore, when traffic load is
low, a low queue threshold should be used to improve energy savings
while maintaining tolerable packet delays. On the contrary, a high
queue threshold should be used at high rates to obtain significant
energy savings. These results demonstrate that a single queue
threshold value does not suit well for every downlink traffic stream,
so an algorithm that dynamically adjusts this parameter in accordance
with existing traffic conditions should be applied.

\subsection{Adaptive Coalesced DRX}

We now evaluate our adaptive mechanism using the same DRX parameters
as in the previous set of experiments. We configured our scheme with
two different target average delays. If the UE were running some delay
tolerant applications, the target delay could be configured with a
high value, so we have firstly conducted several simulation
experiments with $W^*=512\,$ms. On the contrary, in a scenario with
delay sensitive traffic, a low value should be assigned to the target
delay, so we have also conducted some simulations with $W^*=64\,$ms to
evaluate our scheme under these more stringent conditions. Finally, we
set $W_{\max}=2W^*$ in both scenarios.

\begin{figure}[t]
  \centering
  \subfigure[Energy savings.]{
    \includegraphics[width=0.95\columnwidth]{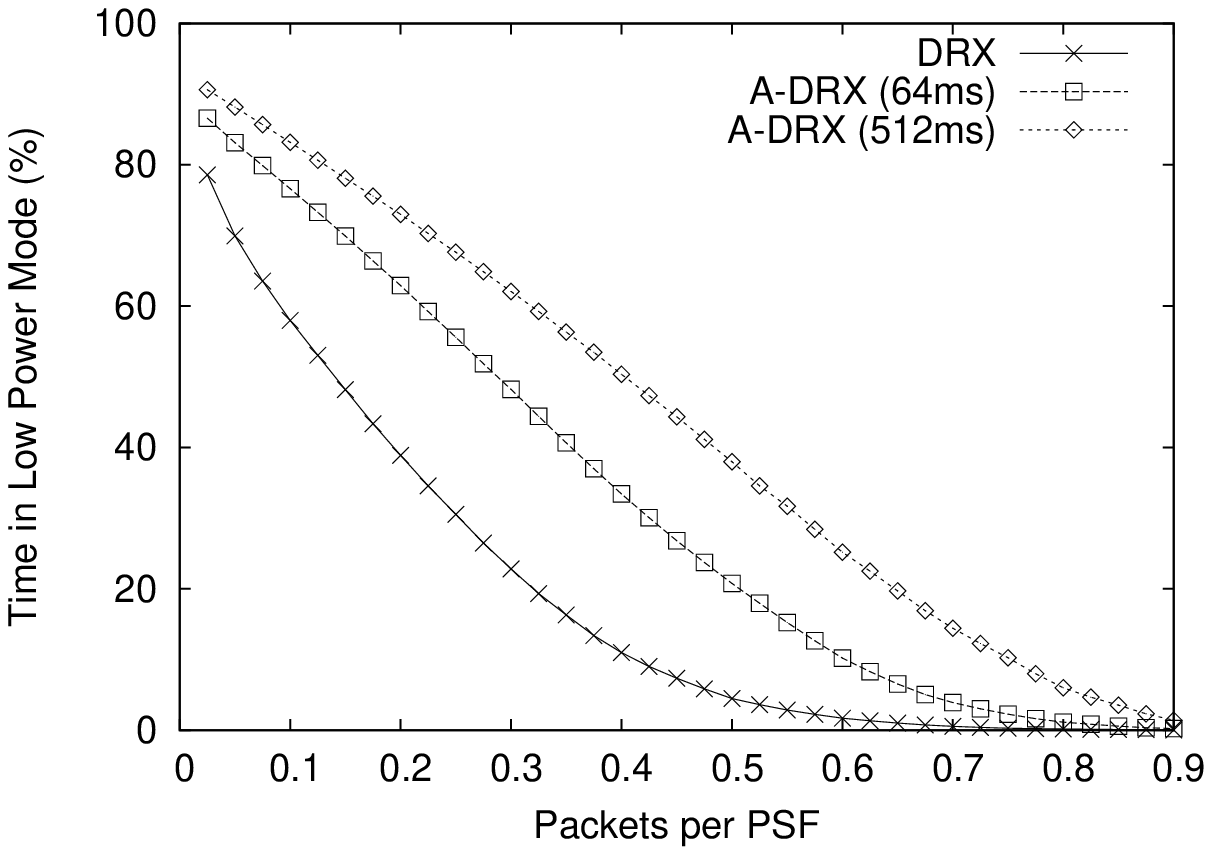}
    \label{fig:adaptive-energy}
  }
  \subfigure[Average queueing delay.]{
    \includegraphics[width=0.95\columnwidth]{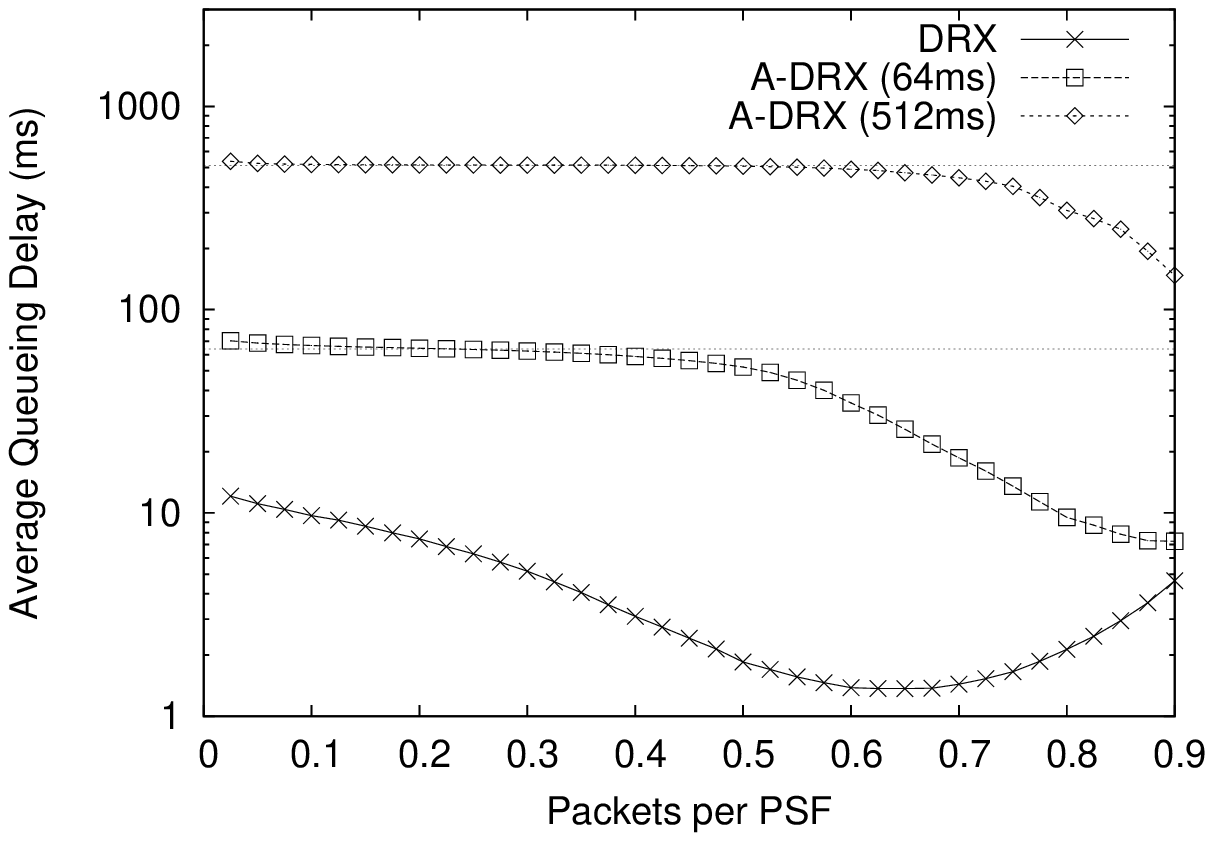}
    \label{fig:adaptive-delay}
  }
  \subfigure[Average queue threshold.]{
    \includegraphics[width=0.95\columnwidth]{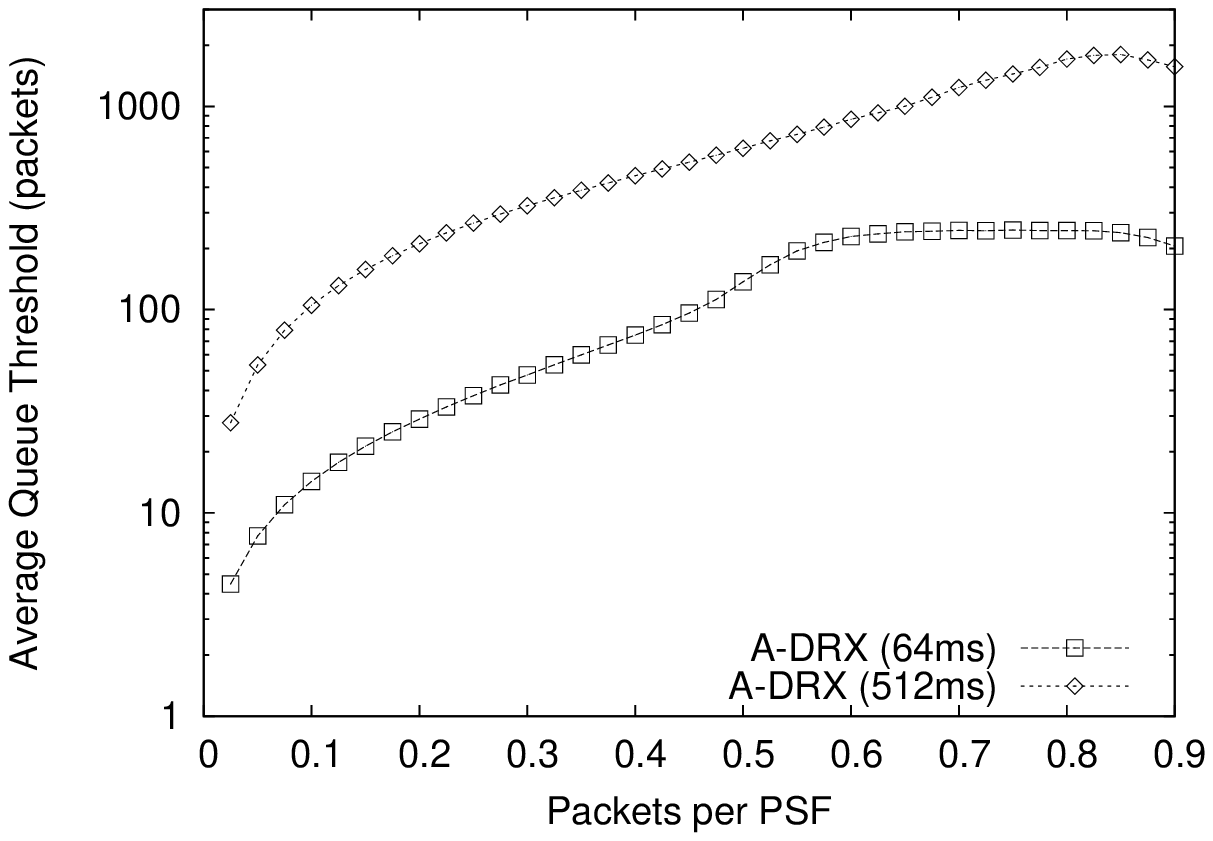}
    \label{fig:adaptive-qth}
  }
  \caption{Results with adaptive coalesced DRX.}
  \label{fig:adaptive}
\end{figure}

Figure~\ref{fig:adaptive} shows the percentage of time spent in the
low power mode and the average queueing delay obtained with
conventional DRX and our adaptive coalesced DRX scheme. As expected,
the adaptive mechanism is able to accommodate the queue threshold to
the traffic load thus achieving significant energy savings while
maintaining, at the same time, the average queueing delay around (or
below) the configured target value. Obviously, the higher $W^*$, the
greater energy savings with larger packet delays we obtained.

Figure~\ref{fig:adaptive-qth} shows that our scheme adjusts the queue
threshold to traffic conditions just choosing higher queue thresholds
as traffic load increases. These higher thresholds enable greater
power savings without sacrificing packet delay, since the time
required to reach them is lower as packet inter-arrival times
decrease.  However, note that, at the highest loads, the average queue
threshold stops increasing and maintains an almost constant
value. This is a consequence of setting a maximum queue
threshold~$Q_{\max}$ to bound the maximum queueing delay. This also
explains why, in Fig.~\ref{fig:adaptive-delay}, the average queueing
delay does not reach the predefined target at high loads.

\subsection{Adaptive Coalesced DRX in Dynamic Scenarios}

In all the previous experiments we simulated static scenarios with
constant arrival rates. With the goal of exploring the speed of
convergence of our algorithm, in the next experiment we simulate for
$100\,$seconds a dynamic scenario in which the arrival rate is updated
every $20\,$seconds following a pattern of $\{0.1,0.2,0.4,0.2,0.1\}$
packets per PSF.

\begin{figure}[t]
  \centering
  \subfigure[Queue threshold.]{
    \includegraphics[width=0.95\columnwidth]{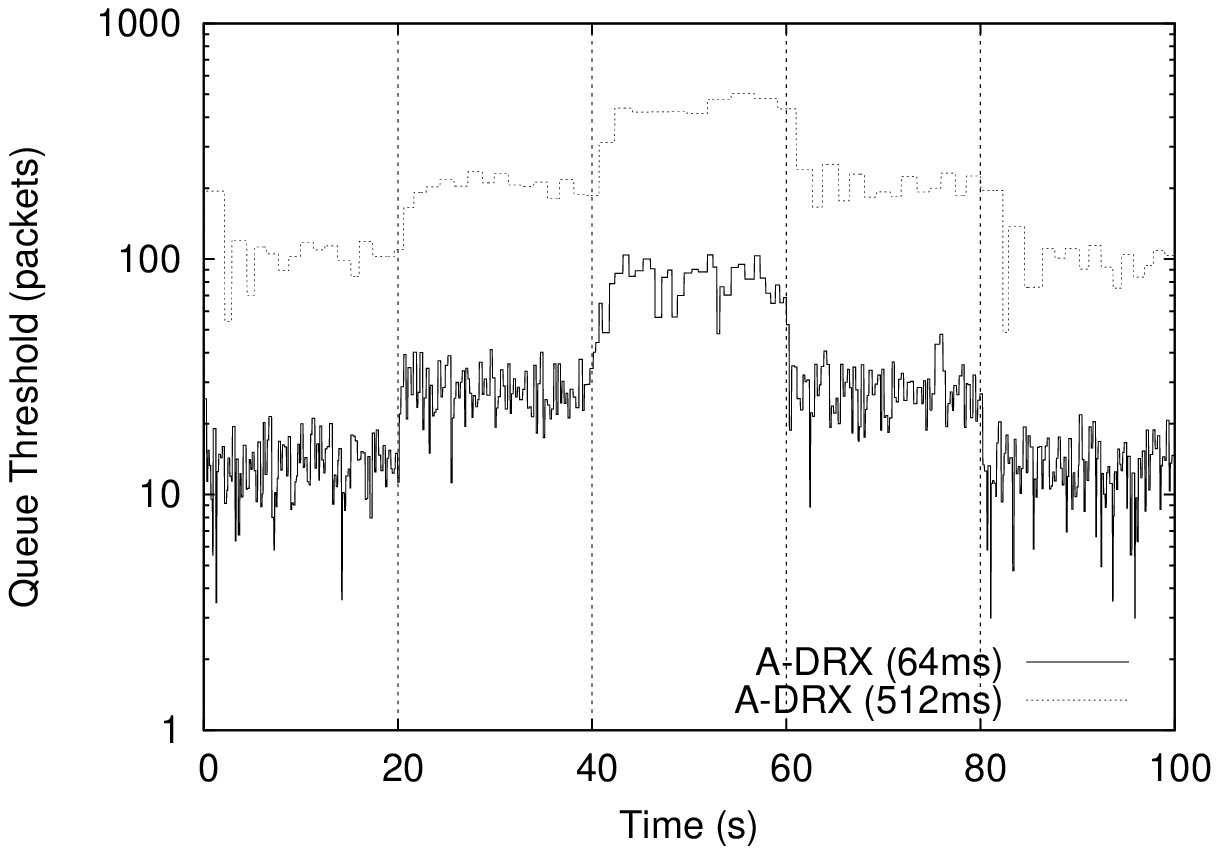}
    \label{fig:adaptive-qth-dyn}
  }
  \subfigure[Energy savings.]{
    \includegraphics[width=0.95\columnwidth]{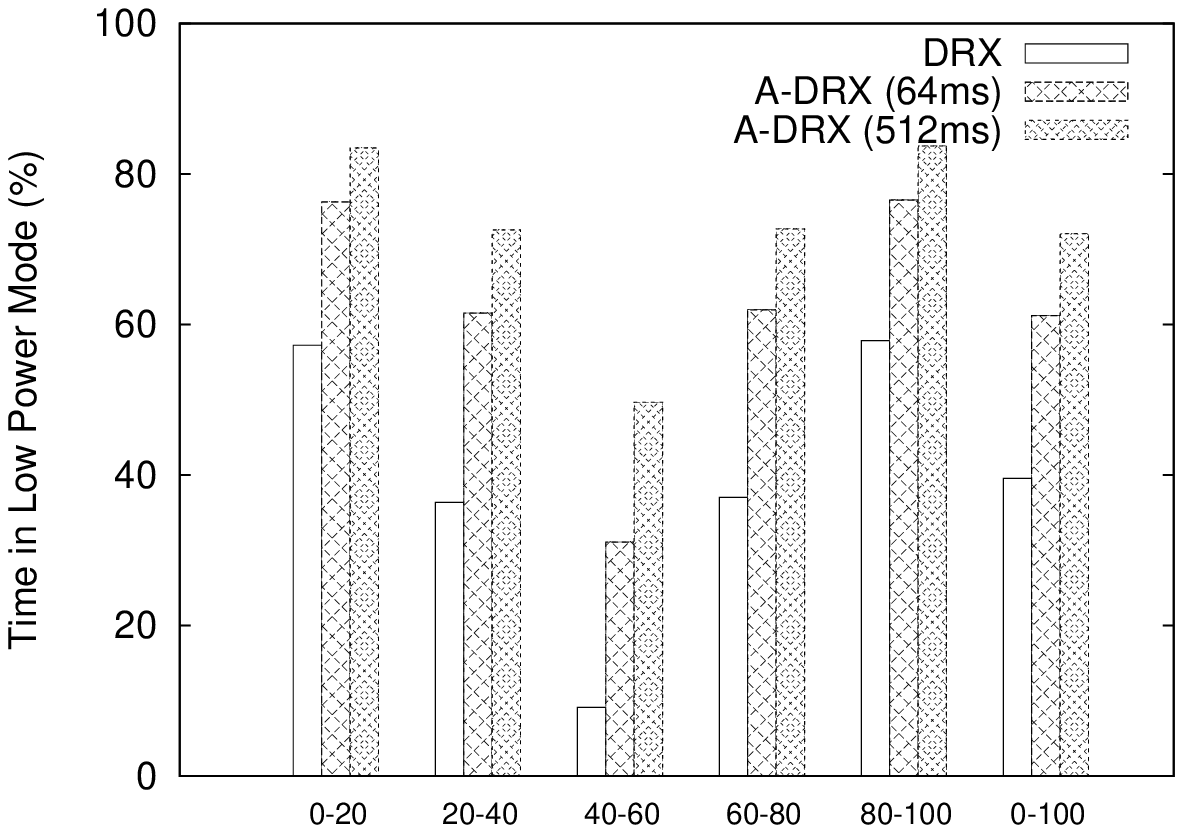}
    \label{fig:adaptive-perc-dyn}
  }
  \subfigure[Average queueing delay.]{
    \includegraphics[width=0.95\columnwidth]{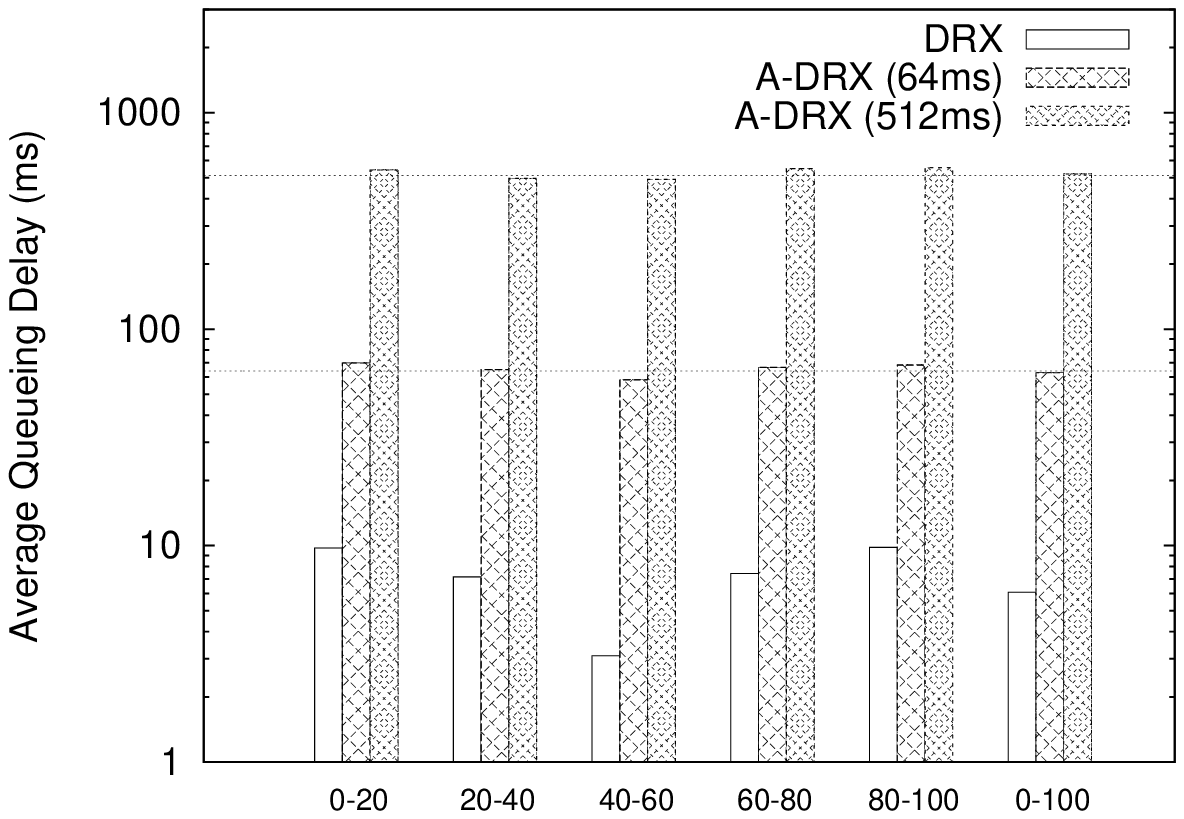}
    \label{fig:adaptive-del-dyn}
  }
  \caption{Results in the dynamic scenario.}
  \label{fig:adaptive-dyn}
\end{figure}

Figure~\ref{fig:adaptive-qth-dyn} shows a representative example of
the evolution of the queue threshold throughout the simulation for
both $W^*$. It can be easily seen how our algorithm is able to quickly
adjust the queue threshold to changing traffic conditions. Also note
that the higher $W^*$ value causes less variations in the queue
threshold since it entails greater queue thresholds and, therefore,
longer coalescing cycles. In Fig.~\ref{fig:adaptive-perc-dyn}
and~\ref{fig:adaptive-del-dyn} we show the energy savings and the
average queueing delay obtained in each interval of $20\,$seconds
respectively. We also show the average over the whole simulated time
in the last cluster of data. As expected, our scheme maintains the
average queueing delay around the target value in all the defined
intervals and, therefore, the time spent in the low power mode is
significantly increased.

\subsection{Adaptive Coalesced DRX with Self-Similar Traffic}

\begin{figure}[t]
  \centering
  \subfigure[Energy savings.]{
    \includegraphics[width=0.95\columnwidth]{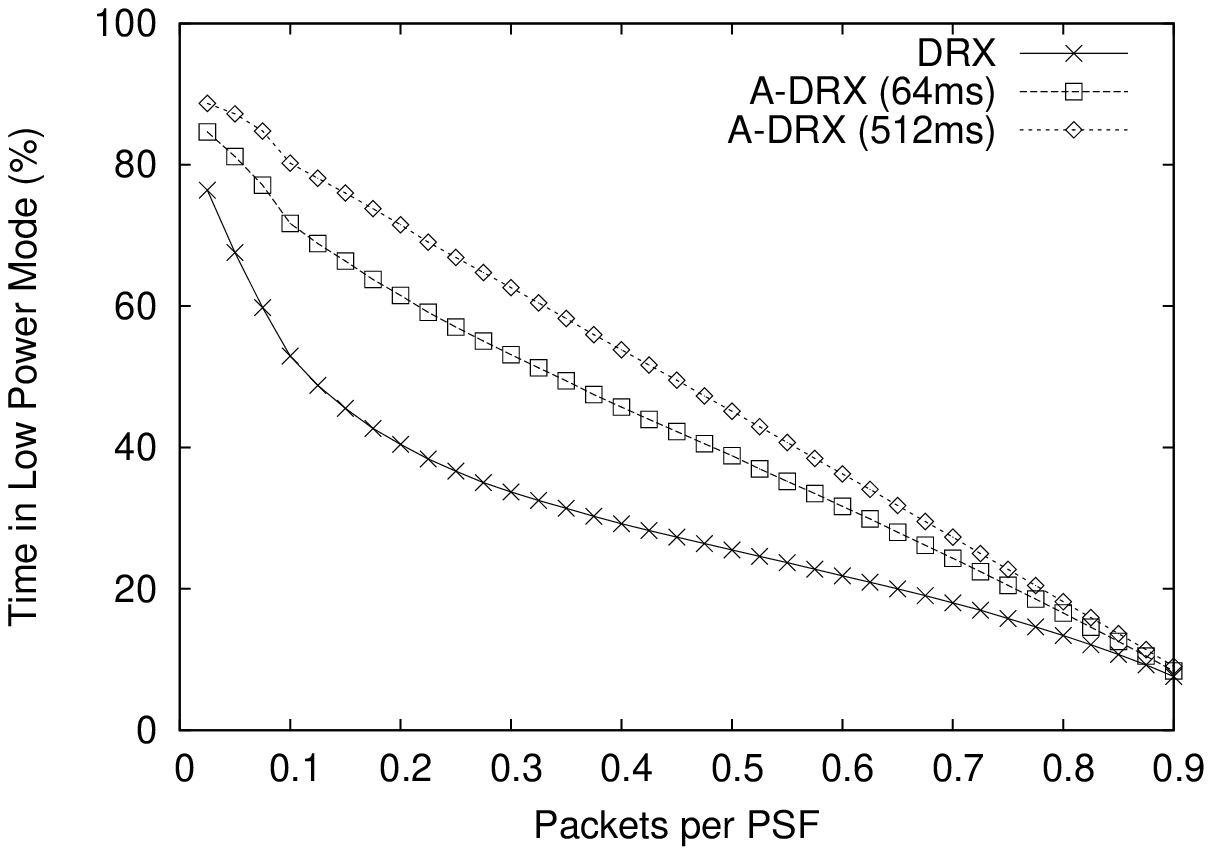}
    \label{fig:pareto-energy}
  }
  \subfigure[Average queueing delay.]{
    \includegraphics[width=0.95\columnwidth]{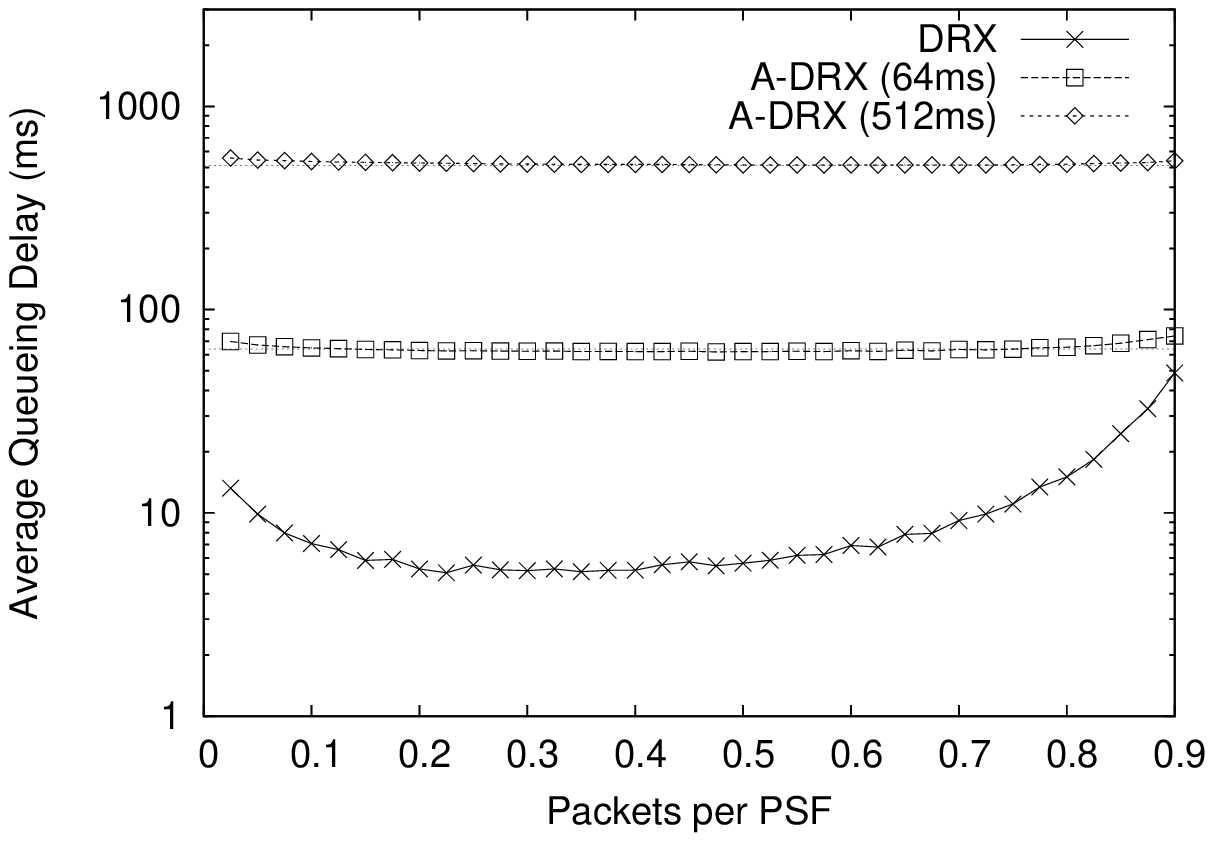}
    \label{fig:pareto-delay}
  }
  \subfigure[Average queue threshold.]{
    \includegraphics[width=0.95\columnwidth]{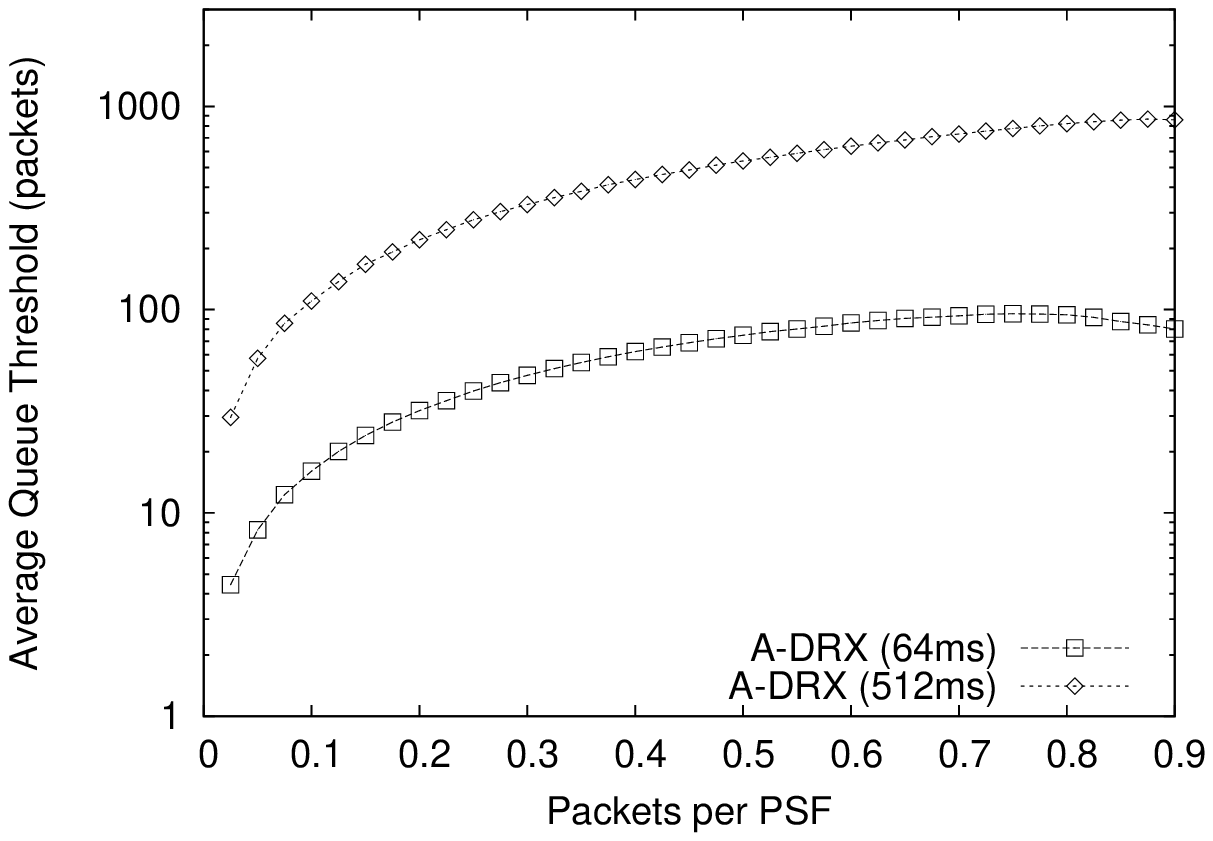}
    \label{fig:pareto-qth}
  }
  \caption{Results with Pareto traffic.}
  \label{fig:pareto}
\end{figure}

We have conducted some extra simulations to test our proposal under
more realistic conditions. For instance, to characterize self-similar
Internet traffic~\cite{crovella97:self_similarity}, in the following
experiments we consider Pareto traffic with shape parameter
$\alpha=1.5$. The results obtained with Pareto traffic are shown in
Fig.~\ref{fig:pareto}. As in the previous experiments, our proposal
can achieve greater energy savings than standard DRX while maintaining
the average queueing delay bounded at the same time. However, note
that our scheme is now able to keep the average queueing delay closer
to the target value even for the highest rates since Pareto traffic
requires lower queue thresholds than Poisson traffic.

\subsection{Adaptive Coalesced DRX with Video Streaming Traffic}

In recent years, mobile networks have experienced a huge increase in
data traffic mainly due to video streaming services. In fact, it is
expected that this trend continues in the next few years and Internet
video services grow to account for more than $50\%$ of mobile data
traffic in 2019 (up from around $40\%$
today)~\cite{ericsson14:mob_report}.

To assure that our proposal remains valid with this important traffic
class, we fed the simulator with some traces from real video streaming
applications previously used
in~\cite{hoque15:mob_multimedia_streaming,eittenberger13:monitor_mob_video}. Traces
from two different applications have been examined: YouTube, which
uses HTTP streaming, and SopCast, which is a popular peer-to-peer
(P2P) live streaming application~\cite{sopcast}. SopCast uses a
proprietary P2P streaming protocol to transmit the video content via
UDP. The traces were collected using their respective native Android
apps from several phones and tablets over different mobile networks.

\begin{figure}[t]
  \centering
  \subfigure[Energy savings.]{
    \includegraphics[width=0.95\columnwidth]{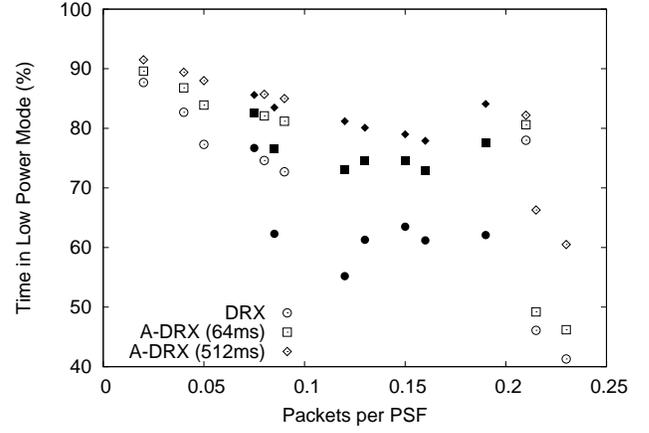}
    \label{fig:traces-energy}
  }
  \subfigure[Average queueing delay.]{
    \includegraphics[width=0.95\columnwidth]{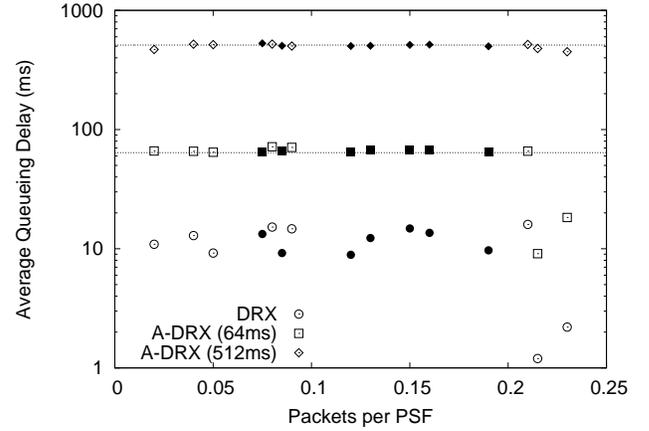}
    \label{fig:traces-delay}
  }
  \caption{Results with video streaming traces. YouTube results are
    shown with unfilled points while filled points are used to show
    the results with SopCast.}
  \label{fig:traces}
\end{figure}

Figure~\ref{fig:traces} shows some representative results obtained
using these real traffic traces. As in the previous experiments with
simulated traffic, our proposal is able to achieve notable
improvements on energy savings, especially with the SopCast
application, while keeping the average queueing delay close to the
target value.

\section{Discussion and Related Work}
\label{sec:related}

Unlike most DRX schemes proposed to improve energy efficiency of UEs
\cite{bontu09:drx,karthik13:pr_alg_drx,koc14:drx_config,wen12:perf_drx,yu12:adj_drx,alouf12:power_saving_analysis,wang14:model_drx,wang14:drx_aware,tseng15:model_drx},
the technique discussed in this paper does not require DRX
reconfiguration and, therefore, does not increase RRC signaling
overhead. Notice that an excess of signaling overhead has a
non-negligible cost on energy efficiency since it adds considerable
processing load to UEs and to other networking
components~\cite{gupta13:eval_sign}. Additionally, these schemes
require to solve a relatively complex optimization problem each time
the traffic conditions at any UE change, thus increasing significantly
the processing load of the eNB. Our proposal, on the contrary, just
requires a few simple computations per UE to be done.

To the best of our knowledge, CDA-DRX is the only proposal able to
reduce energy consumption without introducing extra signaling
overhead~\cite{liu13:adaptive_power_saving}. CDA-DRX allows UEs to
autonomously adjust their DRX cycles according to ongoing user
activity by using two synchronized counters in both the UE and the
eNB. One of these counters keeps account of the number of consecutive
active DRX periods while the other one counts consecutive idle DRX
periods. Two trigger thresholds, one for each of these counters, are
also defined so that, when each counter reaches its corresponding
threshold, the length of the DRX cycle is extended (or reduced)
simultaneously in both the UE and the eNB without the need of DRX
reconfiguration. Unfortunately, this scheme requires modifying the
3GPP RRC protocol~\cite{3gpp_ts_36331} to permit UEs to indicate to
the eNB whether they support CDA-DRX and, if so, to negotiate between
them the involved CDA-DRX parameters (that is, the trigger thresholds
and the series of DRX cycle lengths supported). On the contrary, our
scheme only requires some minor changes to eNB operations, so existing
wireless protocols can still be used without any modification at all.

Finally, we would like to highlight that our scheme just requires the
configuration of two straightforward parameters: the average and the
maximum delay desired for downlink traffic at the eNB. These
parameters could be configured taking into account the power
preference indication (PPI) sent by the UE through the UE assistance
information RRC message~\cite{3gpp_ts_36331}. If the UE is running
delay tolerant applications, it will send the PPI bit active and,
then, the eNB could select high values for the target
delays. Conversely, with delay sensitive traffic, the PPI bit will be
set to zero and the eNB should select lower values for
them.\footnote{For those earlier systems to 3GPP Release 11 that lack
  a way to communicate their power preferences to the network, the eNB
  could configure coalesced DRX parameters without UE assistance
  trying, for example, some deep packet inspection to guess running
  applications and their corresponding delay requirements. Obviously,
  this approach requires adding quite extra overhead at the eNBs, so,
  when possible, the use of the PPI bit to communicate UE preferences
  to the eNB should be preferred.} Maximum target delays around
$50$--$100\,$ms for delay sensitive applications and around $300\,$ms
for delay tolerant applications are suitable~\cite{3gpp_ts_23203}. If
the UE traffic is composed of several flows with different degrees of
delay sensitivity, the target delay should be configured with a value
suitable to the most stringent flow and the eNB should schedule the
transmission of UE packets according to their delay
requirements~\cite{liang13:energy_efficient_scheduling}.

\section{Conclusions}
\label{sec:conclusions}

This paper presents a promising DRX scheme able to improve energy
efficiency of UEs while maintaining the average packet delay bounded
at the same time. Essentially, we propose that eNBs delay downlink
transmission until their downstream queues reach a threshold, thus
increasing the amount of time the UEs spend in DRX mode. Since a
single value for this threshold does not suit well for all possible
traffic loads, we have also presented an adaptive algorithm able to
adjust the queue threshold in accordance with existing traffic
conditions.

Unlike other power saving schemes proposed in the literature, our
mechanism does not increase RRC signaling overhead since it does not
rely on DRX reconfiguration. Furthermore, it can be easily deployed in
LTE/LTE-A~networks since it is very simple to configure and does not
require any changes to the wireless protocols in operation at the
current time.

\section*{Acknowledgments}

The authors thank Dr. Mohammad Ashraful Hoque from the University of
Helsinki for kindly providing us with some of the traffic traces
employed in~\cite{hoque15:mob_multimedia_streaming}. Traces used
in~\cite{eittenberger13:monitor_mob_video} were obtained from the
UMass Trace Repository~\cite{umass_trace_repo}.

Work supported by the European Regional Development Fund (ERDF) and
the Galician Regional Government under agreement for funding the
Atlantic Research Center for Information and Communication
Technologies (AtlantTIC).

\appendix[Stability Analysis]

The adaptive $Q_{\mathrm{w}}$ tuning scheme proposed in
Algorithm~\ref{alg:tuning} can be modeled as the following dynamical
system:
\begin{flalign}
  \label{eq:dynamics}
  &Q_{\mathrm{w}}[i+1] = \nonumber \\
  &\min\left\{\max \left\{ Q_{\mathrm{w}}[i] + 2\lambda (W^*-f(Q_{\mathrm{w}}[i])), 1
  \right\}, Q_{\max} \right\}
\end{flalign}
in the discrete-time index $i=1,2,\dots,$ where $f(\cdot)$ captures
the dependence of the average queueing delay on the queue
threshold. Since $f(\cdot)$ is a continuous, increasing and
differentiable function of $Q_{\mathrm{w}}$ with bounded derivative,
this system clearly reaches the fixed equilibrium point
$Q^*_{\mathrm{w}}$ when $f(Q^*_{\mathrm{w}})=W^*$. Then, assuming $1
\le Q^*_{\mathrm{w}} \le Q_{\max}$ since, otherwise, the target delay
$W^*$ is not achievable, \eqref{eq:dynamics} can be written as
\begin{equation}
  \label{eq:dynamics2}
  Q_{\mathrm{w}}[i+1] = Q_{\mathrm{w}}[i] + 2\lambda (W^*-f(Q_{\mathrm{w}}[i])).
\end{equation}

It is straightforward to prove via linearization that this system is
stable if the derivative of \eqref{eq:dynamics2} at the equilibrium
point has an absolute value strictly less than one, that is,
\begin{equation}
  \label{eq:stability}
  \left| 1 - 2\lambda f'(Q^*_{\mathrm{w}}) \right| < 1.
\end{equation}
Therefore, the stability condition is met if $0 < f'(Q^*_{\mathrm{w}})
< 1/\lambda$. From~\eqref{eq:dW_dQ}, the condition $f'(Q^*_{\mathrm{w}})
< 1/\lambda$ holds if and only if
\begin{equation}
  \label{eq:stab1}
  \frac{\lambda
    T_{\mathrm{w}}-\gamma(\gamma+1)}{(Q^*_{\mathrm{w}}+\lambda
    T_{\mathrm{w}}+\gamma-1)^2} - \frac{2}{{Q^*_{\mathrm{w}}}^2} < 1,
\end{equation}
what is true for all $\lambda$ and $Q^*_{\mathrm{w}} \ge 1$ since
$\gamma \ge 1$ as proved in Section~\ref{sec:gamma}. On the other
hand, the condition $f'(Q^*_{\mathrm{w}}) > 0$ holds if and only if
\begin{equation}
  \label{eq:stab2}
  \frac{2}{{Q^*_{\mathrm{w}}}^2} - \frac{\lambda
    T_{\mathrm{w}}-\gamma(\gamma+1)}{(Q^*_{\mathrm{w}}+\lambda
    T_{\mathrm{w}}+\gamma-1)^2} < 1.
\end{equation}
To check this condition, we must consider two different cases. If
$\lambda T_{\mathrm{w}}-\gamma(\gamma+1) \ge 0$, that is, $\lambda \ge
\gamma(\gamma+1)/T_{\mathrm{w}}$, \eqref{eq:stab2}~is true for all
$Q^*_{\mathrm{w}} > \sqrt{2}$. Clearly, this condition is fulfilled in
almost all reasonable scenarios since it is highly probable that the
equilibrium point surpasses~$\sqrt{2}$ under the previous assumption
of high arrival rates. Unfortunately, $\lambda
T_{\mathrm{w}}-\gamma(\gamma+1) < 0$ with usual DRX parameters. In
this case, the LHS of~\eqref{eq:stab2} goes to zero when
$Q^*_{\mathrm{w}}\to \infty$, so there must exist a value $q$ of the
queue threshold such that \eqref{eq:stab2}~holds for any
$Q^*_{\mathrm{w}}>q$ guaranteeing system stability. Usually, the
$Q^*_{\mathrm{w}}$ required to achieve a given target delay $W^*$ is
high enough to fulfill~\eqref{eq:stab2}. This is the case for all the
simulated scenarios. However, if $W^*$ is too stringent,
$Q^*_{\mathrm{w}}$ may be smaller than~$q$ at the lowest rates, so, in
those rare scenarios with excessively low target delays, the stability
of the algorithm could not be guaranteed.

\bibliographystyle{IEEEtran} 
\bibliography{IEEEabrv,1570182913}

\end{document}

%% file: drx-example.pstex_t
\begin{picture}(0,0)%
\includegraphics{drx-example.pstex}%
\end{picture}%
\setlength{\unitlength}{4144sp}%
\begingroup\makeatletter\ifx\SetFigFont\undefined%
\gdef\SetFigFont#1#2#3#4#5{%
  \reset@font\fontsize{#1}{#2pt}%
  \fontfamily{#3}\fontseries{#4}\fontshape{#5}%
  \selectfont}%
\fi\endgroup%
\begin{picture}(11811,2354)(439,-5093)
\put(5581,-3796){\makebox(0,0)[lb]{\smash{{\SetFigFont{14}{16.8}{\familydefault}{\mddefault}{\updefault}{\color[rgb]{0,0,0}$T_{\mathrm{on}}$}%
}}}}
\put(8461,-5011){\makebox(0,0)[lb]{\smash{{\SetFigFont{14}{16.8}{\familydefault}{\mddefault}{\updefault}{\color[rgb]{0,0,0}$T_{\mathrm{l}}$}%
}}}}
\put(3781,-3796){\makebox(0,0)[lb]{\smash{{\SetFigFont{14}{16.8}{\familydefault}{\mddefault}{\updefault}{\color[rgb]{0,0,0}$T_{\mathrm{on}}$}%
}}}}
\put(4681,-3796){\makebox(0,0)[lb]{\smash{{\SetFigFont{14}{16.8}{\familydefault}{\mddefault}{\updefault}{\color[rgb]{0,0,0}$T_{\mathrm{on}}$}%
}}}}
\put(7381,-3796){\makebox(0,0)[lb]{\smash{{\SetFigFont{14}{16.8}{\familydefault}{\mddefault}{\updefault}{\color[rgb]{0,0,0}$T_{\mathrm{on}}$}%
}}}}
\put(6661,-5011){\makebox(0,0)[lb]{\smash{{\SetFigFont{14}{16.8}{\familydefault}{\mddefault}{\updefault}{\color[rgb]{0,0,0}$T_{\mathrm{l}}$}%
}}}}
\put(3511,-5011){\makebox(0,0)[lb]{\smash{{\SetFigFont{14}{16.8}{\familydefault}{\mddefault}{\updefault}{\color[rgb]{0,0,0}$T_{\mathrm{s}}$}%
}}}}
\put(4411,-5011){\makebox(0,0)[lb]{\smash{{\SetFigFont{14}{16.8}{\familydefault}{\mddefault}{\updefault}{\color[rgb]{0,0,0}$T_{\mathrm{s}}$}%
}}}}
\put(5311,-5011){\makebox(0,0)[lb]{\smash{{\SetFigFont{14}{16.8}{\familydefault}{\mddefault}{\updefault}{\color[rgb]{0,0,0}$T_{\mathrm{s}}$}%
}}}}
\put(2341,-3796){\makebox(0,0)[lb]{\smash{{\SetFigFont{14}{16.8}{\familydefault}{\mddefault}{\updefault}{\color[rgb]{0,0,0}$T_{\mathrm{in}}$}%
}}}}
\end{picture}%

%% file: coalescing-drx.pstex_t
\begin{picture}(0,0)%
\includegraphics{coalescing-drx.pstex}%
\end{picture}%
\setlength{\unitlength}{4144sp}%
\begingroup\makeatletter\ifx\SetFigFont\undefined%
\gdef\SetFigFont#1#2#3#4#5{%
  \reset@font\fontsize{#1}{#2pt}%
  \fontfamily{#3}\fontseries{#4}\fontshape{#5}%
  \selectfont}%
\fi\endgroup%
\begin{picture}(11811,2354)(439,-5093)
\put(5581,-3796){\makebox(0,0)[lb]{\smash{{\SetFigFont{14}{16.8}{\familydefault}{\mddefault}{\updefault}{\color[rgb]{0,0,0}$T_{\mathrm{on}}$}%
}}}}
\put(8461,-5011){\makebox(0,0)[lb]{\smash{{\SetFigFont{14}{16.8}{\familydefault}{\mddefault}{\updefault}{\color[rgb]{0,0,0}$T_{\mathrm{l}}$}%
}}}}
\put(3781,-3796){\makebox(0,0)[lb]{\smash{{\SetFigFont{14}{16.8}{\familydefault}{\mddefault}{\updefault}{\color[rgb]{0,0,0}$T_{\mathrm{on}}$}%
}}}}
\put(4681,-3796){\makebox(0,0)[lb]{\smash{{\SetFigFont{14}{16.8}{\familydefault}{\mddefault}{\updefault}{\color[rgb]{0,0,0}$T_{\mathrm{on}}$}%
}}}}
\put(7381,-3796){\makebox(0,0)[lb]{\smash{{\SetFigFont{14}{16.8}{\familydefault}{\mddefault}{\updefault}{\color[rgb]{0,0,0}$T_{\mathrm{on}}$}%
}}}}
\put(6661,-5011){\makebox(0,0)[lb]{\smash{{\SetFigFont{14}{16.8}{\familydefault}{\mddefault}{\updefault}{\color[rgb]{0,0,0}$T_{\mathrm{l}}$}%
}}}}
\put(3511,-5011){\makebox(0,0)[lb]{\smash{{\SetFigFont{14}{16.8}{\familydefault}{\mddefault}{\updefault}{\color[rgb]{0,0,0}$T_{\mathrm{s}}$}%
}}}}
\put(4411,-5011){\makebox(0,0)[lb]{\smash{{\SetFigFont{14}{16.8}{\familydefault}{\mddefault}{\updefault}{\color[rgb]{0,0,0}$T_{\mathrm{s}}$}%
}}}}
\put(5311,-5011){\makebox(0,0)[lb]{\smash{{\SetFigFont{14}{16.8}{\familydefault}{\mddefault}{\updefault}{\color[rgb]{0,0,0}$T_{\mathrm{s}}$}%
}}}}
\put(2341,-3796){\makebox(0,0)[lb]{\smash{{\SetFigFont{14}{16.8}{\familydefault}{\mddefault}{\updefault}{\color[rgb]{0,0,0}$T_{\mathrm{in}}$}%
}}}}
\end{picture}%

%% file: drx-cycle.pstex_t
\begin{picture}(0,0)%
\includegraphics{drx-cycle.pstex}%
\end{picture}%
\setlength{\unitlength}{4144sp}%
\begingroup\makeatletter\ifx\SetFigFont\undefined%
\gdef\SetFigFont#1#2#3#4#5{%
  \reset@font\fontsize{#1}{#2pt}%
  \fontfamily{#3}\fontseries{#4}\fontshape{#5}%
  \selectfont}%
\fi\endgroup%
\begin{picture}(9494,2373)(204,-5053)
\put(586,-2851){\makebox(0,0)[lb]{\smash{{\SetFigFont{12}{14.4}{\familydefault}{\mddefault}{\updefault}{\color[rgb]{0,0,0}$n$-th pkt}%
}}}}
\put(1531,-4381){\makebox(0,0)[lb]{\smash{{\SetFigFont{12}{14.4}{\familydefault}{\mddefault}{\updefault}{\color[rgb]{0,0,0}$n$-th pkt}%
}}}}
\put(8596,-4381){\makebox(0,0)[lb]{\smash{{\SetFigFont{12}{14.4}{\familydefault}{\mddefault}{\updefault}{\color[rgb]{0,0,0}$n+1$-th pkt}%
}}}}
\put(406,-4381){\makebox(0,0)[lb]{\smash{{\SetFigFont{12}{14.4}{\familydefault}{\mddefault}{\updefault}{\color[rgb]{0,0,0}$n-1$-th pkt}%
}}}}
\put(2836,-4921){\makebox(0,0)[lb]{\smash{{\SetFigFont{12}{14.4}{\familydefault}{\mddefault}{\updefault}{\color[rgb]{0,0,0}$T_{\mathrm{in}}$}%
}}}}
\put(4771,-4921){\makebox(0,0)[lb]{\smash{{\SetFigFont{12}{14.4}{\familydefault}{\mddefault}{\updefault}{\color[rgb]{0,0,0}$T_{\mathrm{s}}$}%
}}}}
\put(7516,-4921){\makebox(0,0)[lb]{\smash{{\SetFigFont{12}{14.4}{\familydefault}{\mddefault}{\updefault}{\color[rgb]{0,0,0}$T_{\mathrm{s}}$}%
}}}}
\put(5829,-4741){\makebox(0,0)[lb]{\smash{{\SetFigFont{12}{14.4}{\familydefault}{\mddefault}{\updefault}{\color[rgb]{0,0,0}$T_{\mathrm{on}}$}%
}}}}
\put(8529,-4741){\makebox(0,0)[lb]{\smash{{\SetFigFont{12}{14.4}{\familydefault}{\mddefault}{\updefault}{\color[rgb]{0,0,0}$T_{\mathrm{on}}$}%
}}}}
\put(4456,-2851){\makebox(0,0)[lb]{\smash{{\SetFigFont{12}{14.4}{\familydefault}{\mddefault}{\updefault}{\color[rgb]{0,0,0}$n+1$-th pkt}%
}}}}
\put(6931,-2851){\makebox(0,0)[lb]{\smash{{\SetFigFont{12}{14.4}{\familydefault}{\mddefault}{\updefault}{\color[rgb]{0,0,0}$n+3$-th pkt}%
}}}}
\put(5626,-2851){\makebox(0,0)[lb]{\smash{{\SetFigFont{12}{14.4}{\familydefault}{\mddefault}{\updefault}{\color[rgb]{0,0,0}$n+2$-th pkt}%
}}}}
\put(3556,-4201){\makebox(0,0)[lb]{\smash{{\SetFigFont{12}{14.4}{\familydefault}{\mddefault}{\updefault}{\color[rgb]{0,0,0}$I$}%
}}}}
\put(5176,-3751){\makebox(0,0)[lb]{\smash{{\SetFigFont{12}{14.4}{\familydefault}{\mddefault}{\updefault}{\color[rgb]{0,0,0}$X_n$}%
}}}}
\put(2881,-3121){\makebox(0,0)[lb]{\smash{{\SetFigFont{12}{14.4}{\familydefault}{\mddefault}{\updefault}{\color[rgb]{0,0,0}$A_n$}%
}}}}
\put(6571,-3121){\makebox(0,0)[lb]{\smash{{\SetFigFont{12}{14.4}{\familydefault}{\mddefault}{\updefault}{\color[rgb]{0,0,0}$A_{n+2}$}%
}}}}
\put(1756,-3751){\makebox(0,0)[lb]{\smash{{\SetFigFont{12}{14.4}{\familydefault}{\mddefault}{\updefault}{\color[rgb]{0,0,0}$S_n$}%
}}}}
\put(1081,-3751){\makebox(0,0)[lb]{\smash{{\SetFigFont{12}{14.4}{\familydefault}{\mddefault}{\updefault}{\color[rgb]{0,0,0}$W_n$}%
}}}}
\put(1486,-3436){\makebox(0,0)[lb]{\smash{{\SetFigFont{12}{14.4}{\familydefault}{\mddefault}{\updefault}{\color[rgb]{0,0,0}$D_n$}%
}}}}
\put(5356,-3121){\makebox(0,0)[lb]{\smash{{\SetFigFont{12}{14.4}{\familydefault}{\mddefault}{\updefault}{\color[rgb]{0,0,0}$A_{n+1}$}%
}}}}
\put(6346,-4201){\makebox(0,0)[lb]{\smash{{\SetFigFont{12}{14.4}{\familydefault}{\mddefault}{\updefault}{\color[rgb]{0,0,0}$W_{n+1}=W_{\mathrm{f}}$}%
}}}}
\end{picture}%